\documentclass[lettersize,journal]{IEEEtran}
\usepackage{amsmath,amsfonts}
\usepackage{algorithmic}
\usepackage{array}
\usepackage[caption=false,font=normalsize,labelfont=sf,textfont=sf]{subfig}
\usepackage{textcomp}
\usepackage{stfloats}
\usepackage{setspace}
\usepackage{url}
\usepackage{verbatim}
\usepackage{graphicx}
\def\BibTeX{{\rm B\kern-.05em{\sc i\kern-.025em b}\kern-.08em
    T\kern-.1667em\lower.7ex\hbox{E}\kern-.125emX}}
\usepackage{balance}
\usepackage[table]{xcolor}
\usepackage{amsmath,amsfonts}
\usepackage{algorithmic}
\usepackage{algorithm}
\usepackage{array}
\usepackage[caption=false,font=normalsize,labelfont=sf,textfont=sf]{subfig}
\usepackage{tabularx}
\usepackage{caption}
\usepackage{textcomp}
\usepackage{stfloats}
\usepackage{verbatim}
\usepackage{booktabs}
\usepackage{xcolor}  
\usepackage{cite}
\usepackage{placeins}
\usepackage{times}
\usepackage{moreverb}
\usepackage{mathrsfs}
\usepackage{bm}
\usepackage{url}
\usepackage{framed}
\usepackage{amsmath}
\usepackage{amssymb}
\usepackage[edges]{forest}

\usepackage{multicol}
\usepackage{amsmath,amssymb} 
\usepackage{times}
\usepackage{epsfig}
\usepackage{tabularx}
\usepackage{array}
\usepackage{amstext}
\usepackage{relsize}
\usepackage{lipsum}
\usepackage{color}
\usepackage{multicol}
\usepackage{longtable}
\usepackage{xtab,booktabs}
\usepackage{marginnote}

\newcommand{\tkderevision}{\color{black}}

\usepackage{multirow}
\usepackage{enumitem}
\usepackage{makecell}
\usepackage{amsmath}
\usepackage{textcomp}
\usepackage{stfloats}
\usepackage{url}
\usepackage{verbatim}
\usepackage{tikz}
\usepackage{balance}
\usepackage[colorlinks, linkcolor=blue, citecolor=blue, urlcolor=blue]{hyperref}
\usepackage{ragged2e}
\usepackage{wrapfig}

\newcommand{\hide}[1]{}

\newcommand{\eg}{\textit{e}.\textit{g}.}

\def\BibTeX{{\rm B\kern-.05em{\sc i\kern-.025em b}\kern-.08emT\kern-.1667em\lower.7ex\hbox{E}\kern-.125emX}}

\usepackage{balance}

\definecolor{hidden-draw}{RGB}{20,68,106}
\definecolor{hidden-pink}{RGB}{234, 131, 121}
\definecolor{lightred}{RGB}{220,92,96}
\definecolor{deepblue}{RGB}{125,174,224}
\definecolor{lightpurp}{RGB}{179,149,189}
\definecolor{lightpurple}{RGB}{130, 132, 131}
\definecolor{lightgray}{gray}{0.9}

\definecolor{hiddenc1}{RGB}{59, 118, 122}
\definecolor{hiddenc2}{RGB}{69,105,144}
\definecolor{hiddenc3}{RGB}{130,130,170}

\definecolor{hid-vae}{RGB}{125,174,224}
\definecolor{hid-gnn}{RGB}{179,149,189}
\definecolor{hid-trans}{RGB}{122, 199,226}
\definecolor{hid-dm}{RGB}{225, 225, 255}
\definecolor{hid-llm}{RGB}{84,190,170}
\definecolor{hid-ssl}{RGB}{176,217,146}
\definecolor{hid-dms}{RGB}{238, 144, 59}

\begin{document}

\title{A Survey on Point-of-Interest Recommendation: Models, Architectures, and Security}

\author{Qianru Zhang\IEEEauthorrefmark{1}, Peng Yang\IEEEauthorrefmark{1}, Junliang Yu, Haixin Wang, Xingwei He, Siu-Ming Yiu\IEEEauthorrefmark{2}, Hongzhi Yin\IEEEauthorrefmark{2}
\IEEEcompsocitemizethanks{
\IEEEcompsocthanksitem
\IEEEauthorrefmark{1}Equal Contribution,\IEEEauthorrefmark{2}Corresponding author.
\IEEEcompsocthanksitem
Q. Zhang, P. Yang, X. He and S.M. Yiu are from the University of Hong Kong. E-mail: \{qrzhang,smyiu\}@cs.hku.hk, stuyangpeng@gmail.com, hexingwei15@gmail.com.
\IEEEcompsocthanksitem
H. Wang is from the University of California, Los Angeles. E-mail: whx@cs.ucla.edu.
\IEEEcompsocthanksitem 
J. Yu and H. Yin work at the University of Queensland. E-mail: jl.yu, h.yin1\}@uq.edu.au.
}
}

\pagestyle{empty}

\maketitle

\begin{abstract}
The widespread adoption of smartphones and Location-Based Social Networks has led to a massive influx of spatio-temporal data, creating unparalleled opportunities for enhancing Point-of-Interest (POI) recommendation systems. These advanced POI systems are crucial for enriching user experiences, enabling personalized interactions, and optimizing decision-making processes in the digital landscape. However, existing surveys tend to focus on traditional approaches and few of them delve into cutting-edge developments, emerging architectures, as well as security considerations in POI recommendations. To address this gap, our survey stands out by offering a comprehensive, up-to-date review of POI recommendation systems, covering advancements in models, architectures, and security aspects. We systematically examine the transition from traditional models to advanced techniques such as large language models. Additionally, we explore the architectural evolution from centralized to decentralized and federated learning systems, highlighting the improvements in scalability and privacy. Furthermore, we address the increasing importance of security, examining potential vulnerabilities and privacy-preserving approaches. Our taxonomy provides a structured overview of the current state of POI recommendation, while we also identify promising directions for future research in this rapidly advancing field.
\end{abstract}

\begin{IEEEkeywords}
Point-of-Interest Recommendation, Recommender Systems, Large Language Models, Federated Learning
\end{IEEEkeywords}

\IEEEdisplaynontitleabstractindextext

\IEEEpeerreviewmaketitle

\section{Introduction}\label{sec:intro}
\thispagestyle{empty}

\IEEEPARstart{T}{he} proliferation of smart devices has fueled the rapid growth of Location-based Social Networks (LBSNs)~\cite{ye2010location,bao2015recommendations,chorley2015personality}, which allow users to share check-ins, reviews, and personal experiences tied to specific locations. These networks, now with billions of users, generate vast amounts of spatio-temporal data~\cite{andrienko2007multimodal,sharafi2022novel,du2023cross,zhang2024survey,zhang2023online,zhang2024billiards,zhang2025efficient,zhang2025efficient,zhang2024graph}, presenting valuable opportunities for personalized Point-of-Interest (POI) recommendations. As a dynamic area in recommendation systems, POI recommendation has gained considerable interest from both users and businesses in recent years. These methods leverage users’ historical check-ins along with multimodal data to suggest personalized destinations~\cite{du2023cross,xu2024mmpoi}. However, the diversity in data size, modality, and user expectations introduces new challenges. These complexities motivate researchers to develop innovative techniques that effectively capture mobility patterns and other relevant features, such as spatial, social, and textual information, to enhance the performance of POI recommendations~\cite{rahmani2022systematic,qian2019spatiotemporal,zhang2024graph}.

POI recommendation research has witnessed significant advancements over the past decade, with researchers continually pushing the boundaries along three dimensions: models, architectures, and security.

\begin{itemize}[leftmargin=*]
\item \emph{Model Evolution: From Traditional to Advanced.} In the early stages, POI recommendation systems primarily relied on latent factor models like Latent Dirichlet Allocation (LDA)~\cite{blei2003latent} and Matrix Factorization (MF)~\cite{mnih2007probabilistic} to model dynamic user behavior~\cite{yin2014lcars,yin2015dynamic}. While these methods provided initial solutions, they were limited in capturing the complex patterns of user-POI interactions. The advent of deep learning marked a transformative shift, with models like Long Short-Term Memory (LSTM) networks~\cite{hochreiter1997long} and the Transformer architecture~\cite{vaswani2017attention} proving far more capable of capturing global-scale features and dynamic sequences of user behavior. Alongside this deep learning revolution, the exploration of Graph Neural Networks (GNNs) emerged as particularly suitable for learning representations in POI recommendations~\cite{kim2021dynaposgnn,wang2021attentive,wang2021adq}. GNNs excel at capturing complex dependencies between users and POIs, allowing for more nuanced recommendations. More recently, the field has witnessed rapid advancements with the integration of cutting-edge techniques like Large Language Models (LLMs)~\cite{chang2024survey}, Diffusion Models (DMs)~\cite{yang2023diffusion,ho2020denoising}, and Self-Supervised Learning (SSL)~\cite{jaiswal2020survey}. These innovations have significantly enhanced recommendation accuracy, allowing systems to better model user preferences.

\item \emph{Architecture Transformations: From Centralized to Decentralized and Beyond.} Initially, POI recommendation systems were predominantly server-side~\cite{long2023decentralized,meng2024poi}, relying on centralized processing to manage data and train models. However, this centralized approach soon faced challenges, particularly with scalability and latency, as the growing demand for real-time recommendations strained system performance. To address these issues, the adoption of edge computing emerged, shifting computation closer to the user's device. This transition improved responsiveness and real-time capabilities by reducing the dependency on cloud infrastructure. Building on this momentum, recent advancements in federated learning~\cite{perifanis2023fedpoirec} have introduced a decentralized model training approach. By distributing training across multiple devices, federated learning not only enhances system scalability but also offers stronger privacy protections by keeping user data local and reducing the risks of centralized data processing.

\item \emph{Security Enhancements: From Vulnerable to Robust and Privacy-Preserving.} Parallel to these architectural improvements, POI recommendation systems initially exhibited significant privacy and security vulnerabilities, as early designs were prone to data breaches and exploitation~\cite{wu2022fedattack,yu2023untargeted,zhang2022pipattack}. As these vulnerabilities were exposed, researchers began to focus on developing more secure solutions. Over time, a variety of privacy-preserving techniques were introduced to protect sensitive user data. Approaches such as differential privacy~\cite{chen2022differential} and federated learning~\cite{perifanis2023fedpoirec} have become central to modern POI recommendation systems, ensuring that while user data is safeguarded, the accuracy and relevance of recommendations are maintained. These techniques have shifted the landscape towards more robust systems capable of balancing both security and performance.
\end{itemize}

While existing surveys~\cite{zhao2016survey,werneck2020survey,islam2022survey,wang2023survey,yin2024survey} have contributed valuable insights into POI recommendation systems, there remains a critical need for a comprehensive review that reflects the rapid developments across POI models, architectures, and security. For instance, while Zhao \textit{et al.}~\cite{zhao2016survey} provided a thorough review of POI recommendation with traditional techniques like matrix factorization, the growing challenges within deep POI recommender systems are not covered. By contrast, Wang \textit{et al.}~\cite{wang2023survey} provided an overview of various POI recommendation methods in the deep learning era, yet they do not delve deeply into the architectural and security challenges that have emerged with the rise of decentralized systems and practical privacy concerns. Similarly, Islam's survey~\cite{islam2022survey} emphasizes the impact of deep learning on POI recommendation, but it overlooks key advancements in federated learning and edge computing, which are increasingly shaping the deployment of these systems. Furthermore, Werneck's survey~\cite{werneck2020survey} provides a detailed account of POI recommendation techniques from 2017 to 2019, offering valuable insights into the evolution of methodologies during that time frame, but it lacks coverage of more recent advancements such as the integration of GNNs and LLMs, which bring about not only powerful user modeling capabilities but also intensive computation and scalability issues. \tkderevision{Recently, Yin \textit{et al.}~\cite{yin2024survey} provided a comprehensive survey on on-device recommender systems, covering aspects such as deployment, inference, training, updating, and the security and privacy of these systems. Consequently, their work does not address key aspects of POI recommendation, such as the foundational models used. As shown in Table~\ref{tab:survey}, there is a noticeable gap in the literature regarding cutting-edge models, architectural evolution, and security considerations.

\begin{table}[!ht]
    \centering
    \caption{Comparison of related surveys.}
    \label{tab:survey}
    \small
    \setlength{\tabcolsep}{2.0mm}{
    \begin{tabular}{c|ccccc}
        \toprule
        Survey & Year & Taxonomy & Models & Architecture & Security \\
        \midrule
        \cite{zhao2016survey} & 2016 & $\times$ & $\checkmark$ & $\times$ & $\times$  \\
        \cite{werneck2020survey} & 2020 & $\checkmark$ & $\checkmark$ & $\times$ & $\times$  \\
        \cite{islam2022survey} & 2022 & $\checkmark$ & $\checkmark$ & $\times$ & $\times$  \\
        \cite{wang2023survey} & 2023 & $\checkmark$ & $\checkmark$ & $\times$ & $\times$  \\
        \tkderevision{\cite{yin2024survey}} & \tkderevision{2024} & \tkderevision{$\checkmark$} & \tkderevision{$\times$} & \tkderevision{$\checkmark$} & \tkderevision{$\checkmark$} \\
        \textbf{Ours} & 2024 & $\checkmark$ & $\checkmark$ & $\checkmark$ & $\checkmark$ \\
        \bottomrule
    \end{tabular}}
\end{table}

With the increasing diversity of data sources, the rise of new models and architectures, and the mounting need for privacy-preserving techniques, it is essential to provide an up-to-date, comprehensive survey that covers these critical aspects. By offering an in-depth taxonomy in this article, we aim not only to provide a holistic understanding of POI recommendation but also to pave the way for future research endeavors to build upon this comprehensive groundwork. Our contributions are summarized as follows:
\begin{itemize}[leftmargin=*]
    \item We undertake an exhaustive and contemporary assessment of models, architectures, and security facets within POI recommendation systems, providing intricate insights into the varied methodologies and technologies.
    \item We not only classify existing research on models, architectures, and security but also introduce a new framework to comprehend and structure these essential elements.
    \item Our study highlights several promising areas for future research in POI recommendation, pointing out key topics that are ready for further exploration, encouraging innovation and inspiring researchers to explore new, untapped areas that could shape the future of POI recommendation technology.
\end{itemize}


The remainder of this paper unfolds as follows: 
Section~\ref{sec:problem} delineates the key concepts within POI recommender systems. Section~\ref{sec:tax} presents illustrations of taxonomy. Section~\ref{sec:models} shows preliminaries of models and studies of them. 
Section~\ref{sec:archi} shows preliminaries of different architectures and existing studies of them. Section~\ref{sec:security} shows preliminaries of security knowledge and existing studies of them. Section~\ref{sec:future} outlines potential future directions. Lastly, Section~\ref{sec:conclu} encapsulates the conclusion, highlighting key takeaways.


\section{Preliminaries}
\label{sec:problem}

The process of POI recommendation presents a series of challenges centered around identifying suitable upcoming POIs for users based on their historical check-in data and other pertinent information within an LBSN. Consider a group denoted as $ U = \left\{u_1, u_2, \ldots, u_N\right\} $, comprising $ N $ users within the LBSN, and a set $ P = \left\{p_1, p_2, \ldots, p_M\right\} $ representing $ M $ POIs. Users form connections through a network denoted as $ \tilde{U} = \left\{(u_i, u_j) | u_i, u_j \in U\right\} $. Each POI $ p $ is defined by its geographical coordinates, specifically latitude $ x_p $ and longitude $ y_p $, along with a set of attributes $ A_p $ describing the POI's semantics. The primary objective is to provide personalized recommendations of relevant POIs based on users' past interactions and preferences within the LBSN. {\tkderevision{Table~\ref{tab:notation} provides a concise description of the primary notations used in this paper.}}

\begin{table}[!ht]  
    \centering
    \small
    \caption{Notations used throughout this paper.}
    \label{tab:notation}
    \begin{tabular}{c|p{0.36\textwidth}}
        \toprule
        \textcolor{black}{Notation} & \textcolor{black}{Description} \\ 
        \midrule
        \textcolor{black}{$u_i$} & \textcolor{black}{User $i$ where $i \in \{1, 2, ..., N\}$} \\
        \textcolor{black}{$U$} & \textcolor{black}{Set of $N$ users within the LBSN, where $ U = \left\{u_1, u_2, \ldots, u_N\right\}$.} \\
        \textcolor{black}{$p_i$} & \textcolor{black}{POI, a specific location recognized for its significance or interest.} \\
        \textcolor{black}{$P$} & \textcolor{black}{Set of $M$ POIs, where $P = \{p_1, p_2, ..., p_N  \}$.} \\
        \textcolor{black}{$\tilde{U}$} & \textcolor{black}{Network of user connections within the LBSN, where $ \tilde{U} = \left\{(u_i, u_j) | u_i, u_j \in U\right\}$.} \\
        \textcolor{black}{$x_p$} & \textcolor{black}{Latitude of POI $p$.} \\
        \textcolor{black}{$y_p$} & \textcolor{black}{Longitude of POI $p$.} \\
        \textcolor{black}{$A_p$} & \textcolor{black}{Set of attributes describing the semantics of POI $p$.} \\
        \textcolor{black}{$c^u_{o}$} & \textcolor{black}{Check-in event where user $u$ declares presence at a location at time $o$.} \\
        \textcolor{black}{$C^u$} & \textcolor{black}{List of check-in events for user $u$, where $ C^u = \left\{ c^u_1, c^u_2, \dots, c^u_k \right\}$.} \\
        \bottomrule
    \end{tabular}
    \vspace{-0.2in}
\end{table}

The key concepts of POI recommendation are as follows:

\begin{itemize}[leftmargin=*]
\item \textbf{Point-of-Interest:} A point of interest, represented as $p_i$, signifies a particular location or site recognized for its significance or interest to various individuals, tourists, or researchers. These points on a map hold specific value or importance and can encompass landmarks, attractions, businesses, historical sites, parks, restaurants, hotels, or any locale of potential interest.
\item \textbf{Check-in Event:} A check-in event, denoted as $c^u_{o}$, involves a user publicly declaring their presence at a venue or location at a specific time step $o$. Typically facilitated through mobile apps or social media, this action shares the user's whereabouts with their social circle. During a check-in, the user selects or searches for a place, such as a restaurant, park, or event space, to indicate their presence.
\item {\tkderevision{\textbf{POI Recommendation:} Given a user $ u $'s check-in list $ C^u = \left\{ c^u_1, c^u_2, \dots, c^u_k \right\} $, where each $ c^u_i $ represents a check-in event at time $ o_i $, the POI recommendation task aims to suggest a set of POIs to the user based on their historical check-ins and preferences.  The goal is to suggest POIs that the user is likely to visit in the future, leveraging their interaction patterns and preferences within the LBSN.}}
\item \textbf{Next POI Recommendation:} Given a user $ u $'s check-in list $ C^u $, the next POI recommendation task revolves around predicting the subsequent POI at time $ o_{i+1}$. It specifically refers to recommending the immediate next POI the user is likely to visit based on their current trajectory and patterns.
\item {\tkderevision{\textbf{Spatial Item Recommendation:}} Given a user's historical interaction data at various spatial locations, the spatial item recommendation task aims to suggest items or products that are contextually relevant to the user based on their location and spatial preferences. The goal is to recommend items that the user is likely to engage with, considering both the spatial distribution of items and the user's past interactions within these geographical regions.}
\end{itemize}

\tikzstyle{my-box}=[
    rectangle,
    draw=hidden-draw,
    rounded corners,
    align=center,
    text opacity=1,
    minimum height=1.5em,
    minimum width=5em,
    inner sep=2pt,
    fill opacity=.8,
    line width=0.8pt,
]

\tikzstyle{leaf-head}=[my-box, minimum height=1.5em,
    draw=black, 
    text=black, font=\normalsize,
    inner xsep=2pt,
    inner ysep=4pt,
    line width=0.8pt,
]

\tikzstyle{leaf-task}=[my-box, minimum height=2.5em,
    draw=black, 
    text=black, font=\normalsize,
    inner xsep=2pt,
    inner ysep=4pt,
    line width=0.8pt,
]
\tikzstyle{leaf-taska}=[my-box, minimum height=2.5em,
    draw=black, 
    text=black, font=\normalsize,
    inner xsep=2pt,
    inner ysep=4pt,
    line width=0.8pt,
]
\tikzstyle{leaf-task1}=[my-box, minimum height=2.5em,
    draw=black, 
    text=black, font=\normalsize,
    inner xsep=2pt,
    inner ysep=4pt,
    line width=0.8pt,
]
\tikzstyle{modelnode-task1}=[my-box, minimum height=1.5em,
    draw=black, 
    text=black, font=\normalsize,
    inner xsep=2pt,
    inner ysep=4pt,
    line width=0.8pt,
]
\tikzstyle{leaf-task2}=[my-box, minimum height=2.5em,
    draw=black, 
    text=black, font=\normalsize,
    inner xsep=2pt,
    inner ysep=4pt,
    line width=0.8pt,
]
\tikzstyle{modelnode-task2}=[my-box, minimum height=1.5em,
    draw=black, 
    text=black, font=\normalsize,
    inner xsep=2pt,
    inner ysep=4pt,
    line width=0.8pt,
]
\tikzstyle{leaf-task3}=[my-box, minimum height=2.5em,
    draw=black, 
    text=black, font=\normalsize,
    inner xsep=2pt,
    inner ysep=4pt,
    line width=0.8pt,
]
\tikzstyle{modelnode-task3}=[my-box, minimum height=1.5em,
    draw=black, 
    text=black, font=\normalsize,
    inner xsep=2pt,
    inner ysep=4pt,
    line width=0.8pt,
]

\tikzstyle{leaf-task4}=[my-box, minimum height=2.5em,
    draw=black, 
    text=black, font=\normalsize,
    inner xsep=2pt,
    inner ysep=4pt,
    line width=0.8pt,
]
\tikzstyle{modelnode-task4}=[my-box, minimum height=1.5em,
    draw=black, 
    text=black, font=\normalsize,
    inner xsep=2pt,
    inner ysep=4pt,
    line width=0.8pt,
]

\tikzstyle{leaf-task5}=[my-box, minimum height=2.5em,
    draw=black, 
    text=black, font=\normalsize,
    inner xsep=2pt,
    inner ysep=4pt,
    line width=0.8pt,
]

\tikzstyle{modelnode-task5}=[my-box, minimum height=1.5em,
    draw=black, 
    text=black, font=\normalsize,
    inner xsep=2pt,
    inner ysep=4pt,
    line width=0.8pt,
]

\tikzstyle{leaf-task6}=[my-box, minimum height=2.5em,
    draw=black, 
    text=black, font=\normalsize,
    inner xsep=2pt,
    inner ysep=4pt,
    line width=0.8pt,
]

\tikzstyle{leaf-task7}=[my-box, minimum height=2.5em,
    draw=black, 
    text=black, font=\normalsize,
    inner xsep=2pt,
    inner ysep=4pt,
    line width=0.8pt,
]
\tikzstyle{leaf-task8}=[my-box, minimum height=2.5em,
    draw=black, 
    text=black, font=\normalsize,
    inner xsep=2pt,
    inner ysep=4pt,
    line width=0.8pt,
]

\tikzstyle{leaf-task9}=[my-box, minimum height=2.5em,
    draw=black, 
    text=black, font=\normalsize,
    inner xsep=2pt,
    inner ysep=4pt,
    line width=0.8pt,
]
\tikzstyle{leaf-task10}=[my-box, minimum height=2.5em,
    draw=black, 
    text=black, font=\normalsize,
    inner xsep=2pt,
    inner ysep=4pt,
    line width=0.8pt,
]

\tikzstyle{modelnode-task6}=[my-box, minimum height=1.5em,
    draw=black, 
    text=black, font=\normalsize,
    inner xsep=2pt,
    inner ysep=4pt,
    line width=0.8pt,
]

\tikzstyle{modelnode-task7}=[my-box, minimum height=1.5em,
    draw=black, 
    text=black, font=\normalsize,
    inner xsep=2pt,
    inner ysep=4pt,
    line width=0.8pt,
]
\tikzstyle{modelnode-task8}=[my-box, minimum height=1.5em,
    draw=black, 
    text=black, font=\normalsize,
    inner xsep=2pt,
    inner ysep=4pt,
    line width=0.8pt,
]
\tikzstyle{modelnode-task9}=[my-box, minimum height=1.5em,
    draw=black, 
    text=black, font=\normalsize,
    inner xsep=2pt,
    inner ysep=4pt,
    line width=0.8pt,
]

\tikzstyle{leaf-paradigms}=[my-box, minimum height=2.5em,
    draw=black, 
    text=black, font=\normalsize,
    inner xsep=2pt,
    inner ysep=4pt,
    line width=0.8pt,
]
\tikzstyle{leaf-others}=[my-box, minimum height=2.5em,
    draw=black, 
    text=black, font=\normalsize,
    inner xsep=2pt,
    inner ysep=4pt,
    line width=0.8pt,
]
\tikzstyle{leaf-other}=[my-box, minimum height=2.5em,
    draw=orange!80, 
    fill=orange!15,  
    text=black, font=\normalsize,
    inner xsep=2pt,
    inner ysep=4pt,
    line width=0.8pt,
]

\tikzstyle{modelnode-task}=[my-box, minimum height=1.5em,
    draw=black, 
    text=black, font=\normalsize,
    inner xsep=2pt,
    inner ysep=4pt,
    line width=0.8pt,
]

\tikzstyle{modelnode-paradigms}=[my-box, minimum height=1.5em,
    draw=black, 
    text=black, font=\normalsize,
    inner xsep=2pt,
    inner ysep=4pt,
    line width=0.8pt,
]
\tikzstyle{modelnode-others}=[my-box, minimum height=1.5em,
    draw=black, 
    text=black, font=\normalsize,
    inner xsep=2pt,
    inner ysep=4pt,
    line width=0.8pt,
]
\tikzstyle{modelnode-other}=[my-box, minimum height=1.5em,
    draw=black, 
    text=black, font=\normalsize,
    inner xsep=2pt,
    inner ysep=4pt,
    line width=0.8pt,
]
\begin{figure*}[!ht]
    \centering
    \resizebox{1\textwidth}{!}{
        \begin{forest}
            forked edges,
            for tree={
                grow=east,
                reversed=true,
                anchor=base west,
                parent anchor=east,
                child anchor=west,
                base=left,
                font=\normalsize,
                rectangle,
                draw=hidden-draw,
                rounded corners,
                align=center,
                minimum width=1em,
                edge+={darkgray, line width=1pt},
                s sep=3pt,
                inner xsep=0pt,
                inner ysep=3pt,
                line width=0.8pt,
                ver/.style={rotate=90, child anchor=north, parent anchor=south, anchor=center},
            }, 
            [
                POI Recommendation,leaf-head, ver
                [
                     Models (\textbf{Sec.~\ref{sec:models}}), leaf-task,text width=8em
                    [
                        Latent Factor Model-Based, leaf-task10, text width=11.5em
                        [LDA-Based, leaf-task8, text width=7.5em[LCARS~\cite{yin2013lcars,yin2014lcars}{, }JIM~\cite{yin2015joint}{, }Geo-SAGE~\cite{wang2015geo}{, }TRM~\cite{yin2015joint}{, }\\LA-LDA~\cite{yin2015modeling}{, }TRM+~\cite{yin2016joint}{, }SPORE~\cite{wang2016spore}{, }ST-LDA~\cite{yin2016adapting}{, }
                        UCGT~\cite{yin2016discovering}{, }\\LSARS~\cite{wang2017location}{, }ST-SAGE~\cite{wang2017st}{, }TPM~\cite{wang2018tpm}, modelnode-task8,text width=29em]]
                        [MF-Based, leaf-task9, text width=7.5em[GeoMF~\cite{lian2014geomf}{, }GeoMF++~\cite{lian2018geomf++}{, }LGLMF~\cite{rahmani2020lglmf}{, }SSTPMF~\cite{davtalab2021poi}{, }\\SQPMF~\cite{wang2024sqpmf}, modelnode-task9,text width=29em]]
                    ]
                    [
                        Classic NN-Based, leaf-taska, text width=9.5em
                            [LSTM-Based, leaf-task1, text width=9.5em
                            [ST-LSTM~\cite{zhang2018discrete}{, }LSTM-S~\cite{zhan2019semantic}{, }PS-LSTM~\cite{zhong2021ps}{, } STMLA~\cite{zhang2022spatio}{, }\\LSMA~\cite{yang2022attention}{, }AT-LSTM~\cite{lai2023poi}{, }ATSD-GRU~\cite{jia2023attention}{, }DLAN~\cite{wu2024dlan},modelnode-task1, text width=29em]
                            ]
                            [Transformer-Based, leaf-task2, text width=9.5em
                            [TLR-M~\cite{halder2021transformer}{, }GETNext~\cite{yang2022getnext}{, }
                            CARAN~\cite{hossain2022caran}{, }JANICP~\cite{zhong2022joint}{, }\\Li~\textit{et al.}~\cite{li2022using}{, }STUIC-SAN~\cite{li2022spatio}{, }AMACF~\cite{zang2021cha}{, }CHA~\cite{wang2022next}{, }\\MVSAN~\cite{li2022multi}{, }HAT~\cite{wu2023reason}{, }
                            STAR-HiT~\cite{xie2023hierarchical}{, }
                            CAFPR~\cite{halder2023capacity}{, }\\
                            FPG~\cite{he2023feature}{, }
                            TGAT~\cite{jiang2023temporal}{, }
                            MobGT~\cite{xu2023revisiting}{, }
                            POIBERT~\cite{ho2022poibert}{, }\\
                            AutoMTN~\cite{qin2022next}{, }
                            CCDSA~\cite{wang2023context}{, }Liu
                            ~\textit{et al.}~\cite{liu2023poi}{, }TDGCN~\cite{cao2023improving}{, }\\
                            TGAT~\cite{jiang2023temporal}{, }
                            MGAN~\cite{wu2023muti}
                            {, }STA-TCN~\cite{ou2023sta}
                            {, }Xia \textit{et al.}~\cite{xia2023effective}{, }\\
                            STTF-Recommender~\cite{xu2023spatio}
                            {, }BayMAN~\cite{xia2023bayes}{, }
                            Kumar \textit{et al.}~\cite{kumar2024modified}{, }
                            \\ImNext~\cite{he2024imnext}
                            {, }BSA-ST-Rec~\cite{cheng2024point}
                            {, }Wu \textit{et al.}
                            ~\cite{wu2024reason}
                            {, }ROTAN~\cite{feng2024rotan},modelnode-task2, text width=29em]
                            ]
                            [GNN-Based, leaf-task3, text width=9.5em
                            [STGCN~\cite{han2020stgcn}{, }Dynaposgnn~\cite{kim2021dynaposgnn}
                            {, }ASGNN~\cite{wang2021attentive}
                            {, }GNN-POI~\cite{zhang2021leveraging}
                            {, }\\ADQ-GNN~\cite{wang2021adq}
                            {, }GN-GCN~\cite{mo2022gn}
                            {, }ATST-GGNN~\cite{li2022attention}
                            {, }Zhao \textit{et al.}\\~\cite{zhao2023poi}
                            {, }HS-GAT~\cite{zhang2024hybrid}
                            {, }PROG-HGNN~\cite{meng2024poi}
                            {, }CGNN-PRRG~\cite{liu2024poi}
                            {, }\\HKGNN~\cite{zhang2024hyper},modelnode-task3, text width=29em]
                            ]
                    ]
                    [
                        SSL-Based Models, leaf-task6, text width=9.5em
                         [SML~\cite{zhou2021self}{, }
                         S2GRec~\cite{li2022self}{, }
                         SSTGL~\cite{liu2023self}{, }
                         SelfTrip~\cite{gao2022self}{, }
                         GSBPL~\cite{wang2023exploring}{, }
                         LSPSL~\cite{jiang2023modeling}{, }\\
                         Gao \textit{et al.}~\cite{gao2023predicting}{, }
                         Wang \textit{et al.}~\cite{wang2024graph}{, }
                         SLS-REC~\cite{fu2024contrastive}{, }
                         CLLP~\cite{zhou2024cllp}{, }, modelnode-task6, text width=40em]
                    ]
                    [
                        Generative, leaf-taska, text width=9.5em
                            [
                            LLM-Based, leaf-task5, text width=9.5em
                            [LLMmove~\cite{feng2024move}{, }Li \textit{et al.}~\cite{li2024large}, modelnode-task5,text width=29em]
                        ]
                        [
                            DM-Based, leaf-task7, text width=9.5em
                            [Diff-POI~\cite{qin2023diffusion}{, }DSDRec~\cite{wang2024dsdrec}
                            {, }Diff-DGMM~\cite{zuo2024diff}
                            {, }DCPR~\cite{long2024diffusion}, modelnode-task7,text width=29em]
                        ]
                            [VAE-Based, leaf-task4, text width=9.5em
                            [WaPOIR~\cite{zhou2023uncertainty}{, }Li \textit{et al.}~\cite{li2022linking}, modelnode-task4, text width=29em]
                            ]
                    ]
                ]
                [
                    Architectures \\(\textbf{Sec.~\ref{sec:archi}}), leaf-paradigms,text width=8em
                    [
                        Centralized, leaf-paradigms, text width=9.5em
                        [
                                GeoIE~\cite{wang2018exploiting}{,} SH-CDL~\cite{yin2017spatial}{,} GETnext~\cite{yang2022getnext}{,} STGN~\cite{zhao2020go}{,} AGRAN~\cite{wang2023adaptive}
                                , modelnode-paradigms, text width=40em]
                    ]
                    [
                        Decentralized, leaf-paradigms, text width=9.5em
                        [
                                MAC~\cite{long2023model}{,} DARD~\cite{zheng2024decentralized}{,} LLRec~\cite{wang2020next}{,} DCLR~\cite{long2023decentralized}, modelnode-paradigms, text width=40em]
                    ]
                    [
                         FL-Based, leaf-paradigms, text width=9.5em
                         [ 
                               FL\&PP-POI~\cite{wang2021poi}{,} Huang~\textit{et al.}~\cite{huang2022geographical}{,}CPF-POI~\cite{ye2023adaptive}{,} SCFL~\cite{zhong2024scfl}{,} SFL~\cite{dong2024sfl}{,} FedPoiRec~\cite{perifanis2023fedpoirec}{,}\\
                               PrefFedPOI~\cite{zhang2023fine}{,} PREFER~\cite{guo2021prefer}{,} Dong~\textit{et al.}~\cite{dong2023sequential}{,} DMF~\cite{chen2018privacy}\text{,}           
                                NRDL~\cite{an2024nrdl}
                         , modelnode-paradigms, text width=40em]
                    ]
                ]
                [
                    Security (\textbf{Sec.~\ref{sec:security}}), leaf-others,text width=8em
                    [
                        {\tkderevision{System Vulnerability}}\\ {\tkderevision{Analysis}}, leaf-others, text width=9.5em
                        [FedRecAttack~\cite{rong2022fedrecattack}\text{,} Ghost riders~\cite{wang2018ghost} \text{,}  LOKI~\cite{zhang2022loki}\text{,} Yuan~\textit{et al.}~\cite{yuan2023manipulating}\text{,} PTIA~\cite{long2024physical}
                                , modelnode-others, text width=40em]
                    ]
                    [
                         User Privacy Protection, leaf-others, text width=9.5em
                         [DMF~\cite{chen2018privacy}\text{,} Li~\textit{et al.}~\cite{kuang2020providing}\text{,} PriRec~\cite{chen2020practical}\text{,} Liu~\textit{et al.}~\cite{liu2017privacy}\text{,} Huo~\textit{et al.}~\cite{huo2021privacy}\text{,} FedPoiRec~\cite{perifanis2023fedpoirec}, modelnode-others, text width=40em]
                    ]
                ]
            ]
        \end{forest}
    }
    \caption{Taxonomy of existing studies for POI recommendation in terms of models, architecture and security}
    \label{fig_tax}
    \vspace{-0.2in}
\end{figure*}
\section{Taxonomy} \label{sec:tax}



This paper distinguishes itself from existing surveys on POI recommendation by adopting a holistic approach that focuses on three critical, interrelated aspects: models, architectures, and security. This tripartite framework serves as a comprehensive taxonomy for analyzing and comparing various POI recommendation systems. 

To contextualize the research landscape, Fig.~\ref{fig_tax} illustrates the framework, while Fig.~\ref{fig:pie_tree} (a) presents the distribution of existing studies across these three categories. Research on POI recommendation models dominates the field, accounting for over 75\% of studies, reflecting ongoing challenges in improving accuracy, personalization, and context-awareness. In contrast, security-related research remains limited, comprising around 8\% of the literature, underscoring gaps in addressing user privacy, adversarial threats, and data integrity in location-based services. Research on architectures, which ensures scalability and efficiency in real-world deployments, occupies the remaining portion, emphasizing its critical yet often overlooked role. To further illustrate the evolution of methodologies in POI recommendation, Fig.~\ref{fig:pie_tree} (b) provides a structured timeline, where key techniques are arranged chronologically from bottom to top. This visualization highlights the shift from traditional models such as LDA and MF to more advanced techniques, including LSTMs, GNNs, transformers, and large language models (LLMs). Additionally, it categorizes architectures into centralized, decentralized, and federated-learning-based approaches, while security-related concerns are mapped to areas such as user privacy protection and system vulnerability analysis. This progression underscores the increasing sophistication and diversification of POI recommendation research over time.

\begin{figure}[!ht]
    \centering
    \begin{tabular}{c c}
    \hspace{-5.5mm}
      \begin{minipage}{0.24\textwidth}
    	\includegraphics[width=\textwidth]{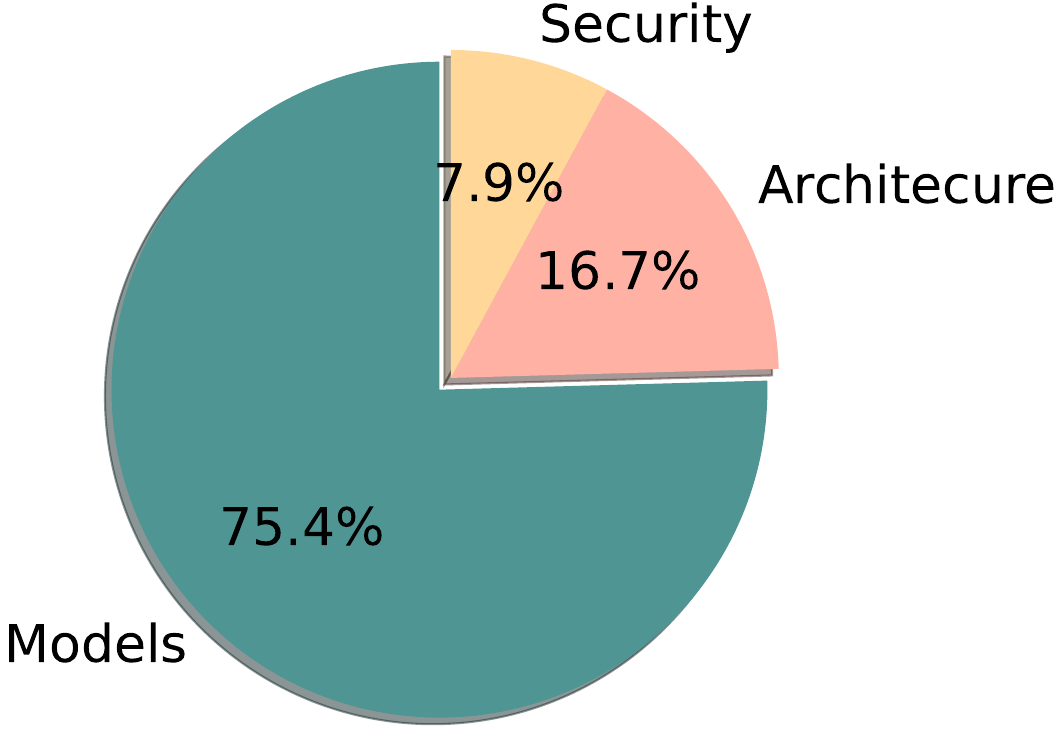}
      \end{minipage}\hspace{-4.00mm}
    &
      \begin{minipage}{0.24\textwidth}
        \includegraphics[width=\textwidth]{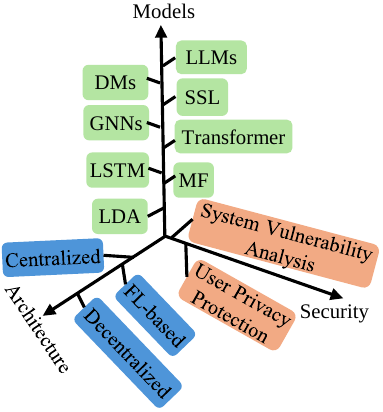}
      \end{minipage}\hspace{-3.0mm}
    \\
    (a) Literature distribution
    &
    (b) {\tkderevision{Development over time}}
    \end{tabular}
    \vspace*{-2mm}
    \caption{Overview of POI recommendation studies.}
    \label{fig:pie_tree}
    \vspace{-0.18in}
\end{figure}


\subsection{Models}

Building on previous POI recommendation surveys~\cite{li2018survey,han2020survey}, which primarily reviewed studies based on traditional non-deep learning and early deep learning methods, we address the advancements brought by more recent techniques. With the rise of novel approaches such as diffusion models, self-supervised learning, and large language models, a growing body of research remains uncovered in existing surveys. To bridge this gap, we provide an extensive review of models from traditional methods to cutting-edge techniques. Detailed explanations of each method, along with the corresponding preliminaries, are provided in Section~\ref{sec:models}.

\subsubsection{Latent Factor Model-Based} 

Traditional POI recommendation methods, such as those based on Latent Dirichlet Allocation (LDA) and Matrix Factorization (MF), have played a foundational role in shaping the field, as noted in prior surveys~\cite{li2018survey,han2020survey}. These approaches generally focus on uncovering latent patterns in user-item interactions to make personalized recommendations. Mathematically, these methods can be formulated as an optimization problem: 
\begin{equation}
\begin{aligned}
\tilde{V} = \arg\min_{\tilde{V}} \mathcal{L}_{f_{\theta}} (U,\tilde{V})
\end{aligned}
\end{equation}
where $f_\theta$ represents the function that minimizes the loss function $\mathcal{L}_{f_{\theta}}$.
$U$ denotes the list of users for whom POI recommendations are being generated.
$\tilde{V}$ represents the list of POIs that minimize the loss function, effectively representing the optimal recommendations.

\subsubsection{Classic Neural Network-Based} 
While latent factor models such as LDA and MF have laid the groundwork for POI recommendation, the advent of deep neural networks, particularly LSTM and Transformer models, has introduced a new paradigm. These models significantly improve both recommendation accuracy and the ability to capture contextual and sequential information. By leveraging the strengths of sequential data modeling and complex feature extraction, neural network-based approaches have made substantial contributions to advancing POI recommendation~\cite{li2023bilstm,halder2021transformer}. These models can be formally represented as follows: 
\begin{equation}
\begin{aligned}
\theta^{\ast} = \arg\min_\theta  - \sum_{i=1}^{U} \sum_{j=1}^{V} y_{ij} \times \log(p_{ij}(\theta))
\end{aligned}
\end{equation}
where $\theta$ denotes the parameters of the neural networks and $U$ is the number of users, $V$ is the number of POIs, $y_{ij}$ represents a binary indicator (1 if user $i$ interacted with POI $j$, 0 otherwise), and $p_{ij}(\theta)$ is the predicted probability of user $i$ interacting with POI $j$, given the model's parameters $\theta$. This equation represents the cross-entropy loss, which is commonly used in POI recommendation. The goal is to minimize the difference between the predicted probabilities and the actual interaction labels. To differentiate this approach from other deep learning-based methods, we categorize it as classic neural network-based, emphasizing its reliance on traditional neural networks, such as LSTM, GNNs, and Transformers.

\subsubsection{Self-supervised Learning-Based}

Self-supervised learning (SSL)~\cite{munro1986self} has gained significant attention in the recommendation domain as a powerful approach to address key challenges like data sparsity and lack of labeled data. Traditional POI recommendation systems often rely on large amounts of user-item interaction data, which may be scarce or incomplete in many cases. SSL provides an alternative by leveraging vast amounts of unlabeled data to create pseudo-supervision signals, enabling the model to learn more effective and generalized representations without needing extensive labeled data~\cite{yu2023self,wang2023exploring,zhou2024cllp}. This makes SSL particularly well-suited for POI recommendation, where interactions are often sparse, and data about user preferences is limited. In the context of POI recommendation, SSL methods often rely on contrastive learning \cite{yu2023self}, which creates tasks where the model learns to differentiate between positive and negative examples. This can be formulated as follows:
\begin{equation}
\begin{aligned}
\mathcal{L}(\theta) =  \sum_{i=1}^{U} \sum_{j=1}^{V}  [d(h(u_i), h(p_j))  -  d(h(u_i), h(p_{j'}))]
\end{aligned}
\end{equation}
where $\theta$ are the parameters of the self-supervised model. $u_i$ is the user embedding for user $i$.
$p_j$ denotes the POI embedding for POI $j$ (positive example).
$p_{j^{\prime}}$ denotes the POI embedding for POI $j^{\prime}$ (negative example).
$h(\cdot)$ denotes the embedding function that maps users and POIs to a shared latent space.
$d(\cdot)$ denotes the distance function (e.g., Euclidean distance) that measures the similarity between embeddings. In this kind of method, the model minimizes the distance between user embeddings and positive POI embeddings while maximizing the distance between user embeddings and negative POI embeddings. It can also be used to pretrain the recommendation model by encouraging the consistency between augmentations of the same user/item while minimizing similarities between augmentations of different users/items.

\subsubsection{Generative}

Generative models use machine learning to discover patterns in data and generate new content. Three commonly used generative models in POI recommendation are large language models (LLMs), diffusion models (DMs), and variational autoencoders (VAEs). LLMs generate coherent text based on input prompts, producing contextually relevant outputs. DMs gradually transform random noise into structured data, creating high-quality images or signals. VAEs encode data into a latent space and decode it to generate new instances.

Large language models, renowned for their ability to comprehend and generate human-like text, have emerged as a transformative force in recommendation systems, including POI recommendations~\cite{feng2024move,li2024large}. Traditionally, recommendation models relied on structured data like user-item interactions, but LLMs introduce a paradigm shift by incorporating rich, unstructured textual data, including user reviews, POI descriptions, and contextual insights such as time, location, and trends. This allows LLMs to exceed traditional recommendation approaches constrained by interaction data. By processing this diverse information, LLMs can generate more personalized, context-aware, and nuanced recommendations. Furthermore, LLMs enhance user experience by engaging users in natural dialogue, making interactions more intuitive. The POI recommendation generation process with LLMs can be formalized as follows:
\begin{equation}
\begin{aligned}
\mathcal{L}(\theta) = - \sum_{j=1}^{U}\sum_{i=1}^{n} \log p(r_i | r_1, ..., r_{i-1}, u_j, c, \theta)
\end{aligned}
\end{equation}
where $\theta$ are parameters of the LLMs and adaptation blocks. $u_j$ is user profile information. $c$ denotes contextual information (e.g., time and location). $r_i$ denotes generated recommendation text. $p(r_i | u_j, c_i, \theta)$ denotes the probability of generating recommendation $r_i$ given user profile $u_j$, context $c$, and parameters $\theta$.

Diffusion models, which learn to gradually add noise to data and reverse the process to generate new samples, can be applied to POI recommendation by generating plausible user-POI interaction patterns. They can create new user-POI interaction data, effectively filling in missing entries in the interaction matrix. This addresses data sparsity and improves recommendation accuracy. This research line can be formulated as follows:
\begin{equation}
\begin{aligned}
\mathcal{L}(\theta) =  E_{x \sim p_{\text{data}}(x)} [D_{KL}(p(x | \theta) || p_{\text{data}}(x))]
\end{aligned}
\end{equation}
where $\theta$ denotes parameters of the diffusion model.
$x$ is user-POI interaction data (e.g., ratings, visits).
$p_{\text{data}(x)}$ is the original data distribution.
$p(x | \theta)$ is the distribution of data generated by the diffusion model with parameters $\theta$.
$D_{KL}(.)$ is Kullback-Leibler divergence, measuring the difference between two probability distributions.

Variational autoencoders~\cite{kingma2013auto,kingma2019introduction} are generative models that learn to represent data in a compressed latent space. They consist of an encoder mapping input data to a distribution in this latent space and a decoder reconstructing data from samples drawn from that distribution. This allows VAEs to generate new instances by sampling from the learned latent space. They enhance POI recommendation systems by capturing user preferences and location characteristics in a compressed space. By encoding user interaction data with various POIs, VAEs capture underlying patterns and preferences, generating personalized recommendations and improving user experience in location-based services.

\subsection{Architectures}

The architecture of a POI recommendation system significantly impacts its ability to handle large volumes of data, ensure privacy and security, and deliver real-time, personalized recommendations. To meet the diverse needs of modern applications and address challenges such as data scalability, user privacy, and computational efficiency, POI recommendation architectures have developed into three primary categories: centralized, decentralized-based, and federated learning-based. Detailed explanations of each architecture, along with the corresponding methods, are provided in Section \ref{sec:archi}.

\subsubsection{Centralized-Based}

In traditional centralized-based architectures, user data and POI information are collected, stored, and processed on a central server. These architectures have long been the standard for POI recommender systems~\cite{yin2017spatial,wang2018exploiting,zhao2018go,wang2023adaptive} due to their ability to aggregate large datasets from multiple users, enabling powerful recommendation algorithms to be applied at scale. However, they face growing concerns related to user privacy, data security, and the cost of managing and scaling centralized infrastructure, particularly as data volumes and user expectations for real-time performance increase. The objective function for optimizing a recommendation model with a centralized architecture can be defined as $\Theta^* = \arg \min_{\Theta} \mathcal{L}(\Theta; \mathcal{D})$
where $\Theta$ denotes the model parameters, and $\mathcal{L}(\Theta; \mathcal{D})$ is the loss function calculated over the entire dataset $\mathcal{D}$. This formulation assumes access to all available data in one place, which, while advantageous for performance optimization, amplifies privacy risks and imposes significant computational overhead.

\subsubsection{Decentralized} 
Decentralized architectures shift the processing burden from centralized servers to users’ individual devices (e.g., smartphones)~\cite{long2023model}. In this approach, recommendation models are deployed locally on the user's device, where data processing and model inference occur without the need to continuously communicate with a central server. This architecture offers several advantages for POI recommendation, such as enhanced privacy, since sensitive user data remains on the device, and reduced latency, as recommendations can be generated in real-time without relying on server-side processing. Decentralized approaches are particularly suitable for mobile or offline scenarios, where data privacy is critical, or network connectivity is limited. However, the performance of decentralized models may be constrained by the device's computational resources, and updating models across devices can be challenging. Formally, the optimization of a decentralized model can be formulated as $\theta^* = \arg \min_{\theta} \mathcal{L}(\theta; \mathcal{D}_u)$, 
where $\theta$ represents the model parameters specific to the device, and $\mathcal{L}(\theta; \mathcal{D}_u)$ is the loss function computed over the local dataset $\mathcal{D}_u$ belonging to the user. By leveraging local computation, decentralized architectures provide a privacy-preserving solution that enables real-time, personalized recommendations tailored to individual user data.

\subsubsection{Federated Learning-Based}

Federated learning (FL)-based architectures~\cite{mcmahan2017communication} represent a newer paradigm that combines the strengths of both centralized and decentralized approaches. In FL, models are trained collaboratively across multiple user devices, where each device computes updates locally using its own data. These updates are then aggregated by a central server to improve a global model without exposing individual user data. This architecture offers a compelling solution for privacy-preserving POI recommendation, as it keeps user data localized while still benefiting from collective learning. FL also reduces the need for direct data transmission between users and servers, addressing privacy concerns and regulatory requirements (e.g., GDPR). Additionally, it offers scalability and adaptability, allowing models to continuously improve as new data is generated on user devices. However, FL introduces challenges in terms of communication overhead, synchronization, and managing heterogeneous data across devices. In the FL-based architecture, the optimization objective of the global model is defined as $\Theta^* = \arg \min_{\Theta} \sum_{k=1}^{K} w_k \mathcal{L}_k(\Theta)$
where $\Theta$ represents the global model parameters, $K$ is the total number of participating clients, $w_k$ is the weight for the $k$-th client, and $\mathcal{L}_k(\Theta)$ is the loss function at the $k$-th client.

\subsection{Security}

Security is crucial in designing POI recommendation systems, especially due to the sensitive nature of location data. Recent work has focused on enhancing user privacy and ensuring system security. Techniques like homomorphic encryption and differential privacy are being developed to protect user data while maintaining recommendation accuracy. Additionally, privacy-preserving protocols are implemented to guard against attacks, including tampering with recommendations or misusing user data. These security efforts are closely tied to the models and architectures used in POI systems, ensuring that they provide accurate recommendations while safeguarding user privacy. A secure POI recommendation system must integrate privacy-preserving techniques at both the model and architecture levels, mitigating security threats without compromising performance. Section~\ref{sec:security} examines the latest research on security in POI recommendations, offering insights into how these considerations are being addressed in practice.


\subsubsection{Data Integrity Threats}

Data integrity threats refer to malicious activities aimed at compromising the accuracy, consistency, and trustworthiness of data within a system. These threats often target the integrity of data at various stages, such as during storage, transmission, or processing. In the context of POI recommendations, data integrity threats can lead to significant issues, such as degraded model performance, biased outputs, and overall system unreliability.

One prominent example of a data integrity threat is poisoning attacks~\cite{rong2022fedrecattack,zhang2022loki}, which manipulate the training data in POI recommendation systems. These attacks can significantly affect the reliability and trustworthiness of the system by injecting malicious data that leads to biased outcomes. Formally, the goal of a poisoning attack can be formulated as:
\begin{equation}
\begin{aligned}
    \tilde{\mathcal{D}} = \arg \max_{\mathcal{D}'} \mathcal{L}_{\text{attack}}(\Theta; \mathcal{D} \cup \mathcal{D}')
\end{aligned}
\end{equation}
where $\tilde{\mathcal{D}}$ represents the manipulated dataset, $\mathcal{L}_{\text{attack}}(\Theta; \mathcal{D} \cup \mathcal{D}')$ is the attack loss function, $\Theta$ are the model parameters, and $\mathcal{D}'$ is the set of malicious data instances added to the original dataset $\mathcal{D}$.

Additionally, POI recommendation systems may face various other threats, including evasion attacks, where adversaries manipulate input data to deceive the system, and Sybil attacks~\cite{wang2018ghost}, where multiple fake profiles are created to influence recommendations. The physical trajectory inference attack~\cite{long2024physical} involves deducing users' historical trajectories from shared data, potentially compromising their privacy by revealing sensitive location information. To defend against such attacks, robust training methods and anomaly detection techniques are employed to identify and mitigate the impact of poisoned data, ensuring the integrity and performance of the recommendation system.

\subsubsection{User Privacy Protection}

In POI recommendation systems, safeguarding user privacy is paramount, as these systems often handle sensitive personal data. Privacy-enhancing technologies play a crucial role in ensuring that user information remains secure while still enabling meaningful data analysis. Techniques such as decentralized secure computation~\cite{chen2018privacy,chen2020practical}, noise injection~\cite{kuang2020providing}, and homomorphic encryption~\cite{liu2017privacy,perifanis2023fedpoirec} are essential in this regard.

For example, homomorphic encryption enables computations on encrypted data, ensuring that raw data remains inaccessible to unauthorized parties. The process can be mathematically expressed as:
\begin{equation}
\begin{aligned}
    \mathbf{Enc}(x) \oplus \mathbf{Enc}(y) &= \mathbf{Enc} (x + y) \\
    \mathbf{Enc}(x) \otimes \mathbf{Enc}(y) &= \mathbf{Enc} (x \cdot y) 
\end{aligned}
\end{equation}
where $\mathbf{Enc}(\cdot)$ denotes the encryption algorithm under a specific key, $\oplus$ and $\otimes$ represent addition and multiplication over ciphertext, $+$ and $\cdot$ represent addition and multiplication over plaintext, and $x, y$ are the input data. This property allows computations to be performed on encrypted data, producing encrypted results that can be decrypted to obtain the final output. These methods facilitate data analysis without compromising the confidentiality of personal information, allowing systems to generate accurate recommendations while maintaining user trust and privacy.

\section{Models}
\label{sec:models}

This section provides a preliminary overview of the \textbf{models} relevant to our survey. We also review specific studies in terms of published venues, techniques, sub-tasks, and datasets in Table~\ref{tab:gen_method}. 

\vspace{-0.1in}
\subsection{Latent Factor Model-Based}

\subsubsection{Latent Dirichlet Allocation}
Early studies in POI recommendation leveraged the power of topic modeling, specifically latent Dirichlet allocation (LDA)~\cite{blei2003latent}, to uncover hidden thematic structures within user activity data. LDA, a generative probabilistic model, assumes that POIs are composed of a mixture of latent topics, each characterized by a distribution over words. This approach, while effective in identifying general themes, often lacked the granularity to capture individual preferences and the dynamic nature of user behaviors.

To address these limitations, researchers began exploring more personalized and context-aware approaches~\cite{wang2017location,wang2018tpm,yin2016discovering,wang2017st,yin2016joint,wang2016spore,xie2016learning,yin2016adapting,yin2015modeling,yin2014lcars,yin2015joint,wang2015geo}. A pioneering work by Yin \textit{et al.}~\cite{yin2014lcars} introduced LCARS, a location-content-aware POI recommender system that combined individual user preferences with the distinct characteristics of a location. This marked a significant shift from purely topic-based models by explicitly incorporating the influence of local trends and preferences on user choices. Building upon this foundation, Wang \textit{et al.}~\cite{wang2018tpm} proposed the Temporal Personalized Model (TPM), which further enhanced POI recommendation by incorporating temporal dynamics and user-specific behavior patterns. TPM introduces the novel concept of topic-regions, clustering locations into semantically meaningful groups based on user activity.


\subsubsection{Matrix Factorization}

Matrix factorization (MF)~\cite{mnih2007probabilistic} is a powerful technique for POI recommendation, addressing sparse user-item interaction data by decomposing it into low-dimensional matrices representing latent user and POI features. This reveals hidden patterns and preferences, enabling personalized recommendations by predicting user affinity for unvisited POIs. Popular methods~\cite{xie2016learning,qian2019spatiotemporal,lian2014geomf,lian2018geomf++} include SVD~\cite{sadek2012svd}, PMF~\cite{brown2015methods}, and NMF~\cite{song2013hierarchical}. Specifically, users tend to exhibit spatially clustered mobility patterns, with visits concentrated within specific geographical regions. This ``clustering phenomenon'' has proven valuable in enhancing POI recommendations, prompting its integration into factorization models. To leverage this insight, Lian \textit{et al.}~\cite{lian2014geomf} propose that user representations are enriched with ``activity area vectors" capturing their typical movement zones, while POI representations incorporate ``influence area vectors" reflecting the regions from which they draw visitors. This fusion of spatial clustering with factorization models allows for a more context-rich and accurate representation of user preferences and POI characteristics, ultimately leading to more effective personalized POI recommendations.

\vspace{-0.1in}
\subsection{Classic Neural Network-Based}

\subsubsection{Long Short-Term Memory}
The long short-term memory (LSTM) network~\cite{hochreiter1997long} is a specialized type of sequential neural network that allows information to persist over time. It is an enhancement of the recurrent neural network (RNN)~\cite{schuster1997bidirectional}, specifically designed to address the vanishing gradient problem, a common issue in RNNs. In the original LSTM architecture~\cite{hochreiter1997long}, at each time step $t$, the input $x_t$ is combined with the previous hidden state $h_{t-1}$. This combined vector is then processed through three key components: the input gate $\mathsf{input}_t$, the output gate $\mathsf{output}_t$, and the input node $\mathsf{gate}_t$, defined by the following equations:
\begin{equation}
\begin{aligned}
    \mathsf{input}_t &= \phi (\mathbf{W}^{(i)} x_t + \mathbf{U}^{(i)} h_{t-1} + b^{(i)})  \\
    \mathsf{output}_t &= \phi (\mathbf{W}^{(o)} x_t + \mathbf{U}^{(o)} h_{t-1} + b^{(o)})  \\
    \mathsf{gate}_t &= \mathsf{tanh} (\mathbf{W}^{(g)} x_t + \mathbf{U}^{(g)} h_{t-1} + b^{(g)}) = \widetilde{C}_t
\end{aligned}
\end{equation}
where $\mathbf{W}^{(*)}$ and $\mathbf{U}^{(*)}$ are weight matrices, $b^{(*)}$ are bias terms, $\widetilde{C}_t$ is the candidate cell state, and $\phi$ is the activation function.

LSTM networks have become indispensable for POI recommendation~\cite{zhang2018discrete,zhao2018go,zhan2019semantic,zhang2022spatio,yang2022attention}, effectively modeling the sequential dependencies inherent in user check-in data to predict future preferences. Earlier studies~\cite{zhang2018discrete,zhao2018go,zhan2019semantic} adopt LSTMs to model POI sequences. Zhao \textit{et al.}~\cite{zhao2018go} address the limitations of traditional RNNs in capturing the crucial influence of time and distance between consecutive check-ins. Their proposed STLSTM model integrates spatio-temporal context, short-term spatio-temporal gating, and long-term spatio-temporal gating. Another study by Zhan \textit{et al.}~\cite{zhan2019semantic} incorporates temporal information to understand the semantic relationships between user actions. They leverage a semantic correlational graph to calculate semantic sequential correlation, factoring in time intervals, and integrate this information into a modified LSTM framework with two additional semantic gates. This enhanced model captures users' sequential behaviors and long- and short-term interests within a semantic context. It also introduces user clustering to improve recommendation accuracy by grouping users based on their semantic preferences.

Recent studies~\cite{yang2022attention,wu2024dlan} based on LSTM mainly aim to capture dynamic dependencies in long-term POI sequences. Yang \textit{et al.}~\cite{yang2022attention} tackle the challenge of effectively capturing the dynamic interplay between a user's long-term preferences and their more immediate, context-dependent interests with AMACF. It achieves this by dynamically weighting input sequences and seamlessly fusing long- and short-term signals. DLAN~\cite{wu2024dlan} incorporates a multi-head attention module that effectively combines first-order and higher-order neighborhood information within user check-in trajectories. This parallel approach addresses the limitations of RNN-based methods, which struggle to establish long-term dependencies between sequences. Furthermore, DLAN integrates a user similarity weighting layer to quantify the influence of social relationships between users on the target user's preferences, leveraging social connections to improve the accuracy of recommendations.

Beyond these studies, others such as Zhong \textit{et al.}~\cite{zhong2021ps} and Zhang \textit{et al.}~\cite{zhang2022spatio} explore innovative ways to leverage LSTMs for POI recommendation, further demonstrating the versatility and effectiveness of this approach. These studies highlight the ongoing evolution of LSTM-based methods for POI recommendation, as researchers continue to develop sophisticated techniques to capture the complexities of user behavior, ultimately leading to more accurate, personalized, and context-aware recommendations.

\subsubsection{Transformer}

The Transformer architecture~\cite{vaswani2017attention}, originally designed for natural language processing, is a deep neural network model that heavily relies on the self-attention mechanism. The architecture comprises two main components: an encoder and a decoder, each consisting of multiple identical layers known as transformer blocks. The encoder's primary function is to process input data and generate corresponding encodings, while the decoder utilizes these encodings, along with contextual information, to generate the final output sequence. As shown in Fig.~\ref{figure-transformer-architecture}, each transformer block includes several key components: a multi-head attention mechanism, a feed-forward neural network and residual connections~\cite{han2020survey}.

\begin{figure*}[h]
\centering
\includegraphics[width=0.70\textwidth]{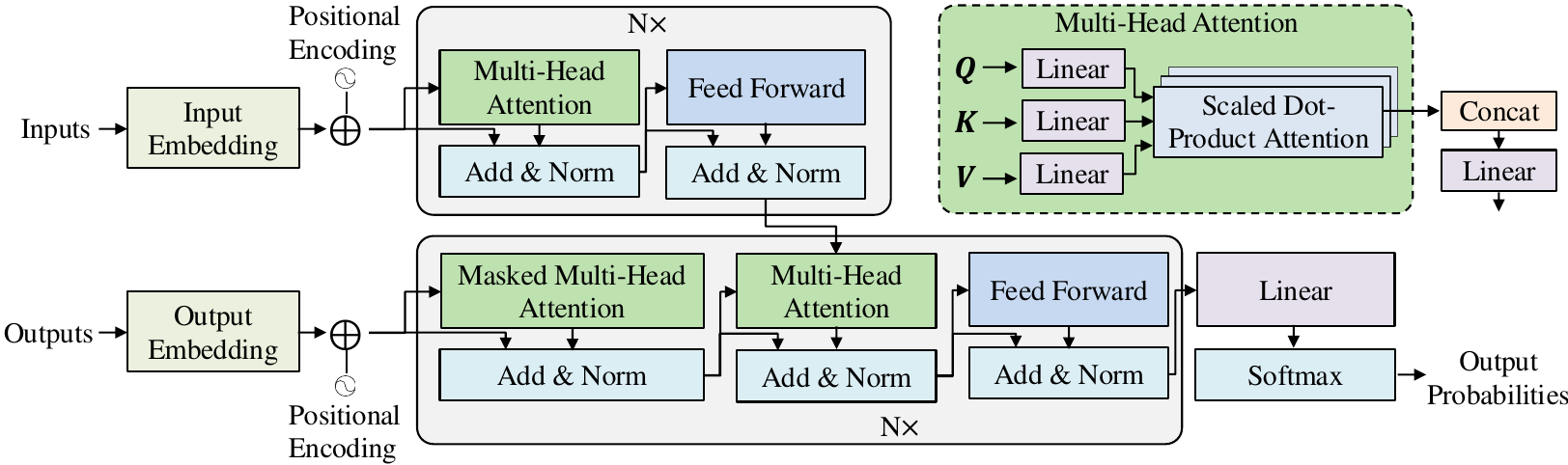}
\caption{The Transformer model architecture}
\label{figure-transformer-architecture}
\vspace{-0.2in}
\end{figure*}

Transformers have revolutionized POI recommendation systems~\cite{halder2021transformer,yang2022getnext,hossain2022caran,zhong2022joint,li2022using,jiang2023temporal}, surpassing traditional methods by effectively capturing complex sequential patterns and contextual information crucial for understanding user preferences. A former study~\cite{halder2021transformer} addresses the challenge of providing queue-time-aware POI recommendations. This task is non-trivial, as it requires both recommending the next POI and accurately predicting the queue time at that location. To tackle this challenge, Halder~\textit{et al.}~\cite{halder2021transformer} propose TLR-M, a multi-task, multi-head attention transformer model. TLR-M simultaneously recommends POIs to target users and predicts the queue time for accessing those POIs. The model's multi-head attention mechanism efficiently integrates long-range dependencies between any two POI visits, enabling it to assess their influence on POI selection.

A recent study~\cite{yang2022getnext} addresses the limitations of treating POI recommendation solely as a sequence prediction task by incorporating collaborative signals through the proposed method GETNext. GETNext introduces a user-agnostic map to capture global movement patterns and collaborative information. It also integrates the flow map into the transformer architecture, enabling the model to learn from both individual user sequences and collective user behavior. Meanwhile, leveraging collaborative signals proves particularly beneficial for recommending POIs to new users with limited historical data. Explicit user preference integration~\cite{li2022using} focuses on directly incorporating explicit user preference data into the transformer model, contrasting with other methods that rely solely on implicit preference learning from POI sequences. It explicitly incorporates data reflecting user preferences, such as ratings or reviews, into the model's input. Directly modeling user preferences aims to provide a more nuanced and accurate representation compared to solely inferring preferences from interaction patterns. The Temporal-Geographical Attention-based Transformer~\cite{jiang2023temporal} tackles the challenge of effectively integrating multiple contextual factors throughout the POI recommendation process. Unlike previous studies, TGAT dynamically selects POI sequences from multiple contextual factor POI graphs derived from user check-in histories to obtain better user representations. Then, TGAT incorporates diverse contextual factors, such as time of day, user interests, and geographical influences, throughout the representation learning process. TGAT also integrates contextual factors, leading to more holistic and accurate representations of POIs and users, thereby enhancing recommendation accuracy.


\subsubsection{Graph Neural Networks}

Graph neural networks (GNNs)~\cite{scarselli2008graph} provide a powerful framework for learning representations from graph-structured data. GNNs operate through a message-passing mechanism, where each node updates its representation by aggregating information from its neighboring nodes and then combining these aggregated messages with its current representation. An illustration of the GNN architecture is presented in Fig.~\ref{fig:model_stru} (a). Consider a graph $G = (V, E)$, where $V$ is the set of nodes and $E$ is the set of edges. Each node $v \in V$ is associated with a feature vector $h'_v$. GNNs utilize the structure of the graph along with these node features $h'_v$ to learn a representation vector for each node, $h_v$, or for the entire graph, $h_G$. Modern GNN architectures typically employ a neighborhood aggregation strategy, where a node’s representation is iteratively refined by aggregating the representations of its neighbors~\cite{xu2018powerful}. After $k$ iterations of aggregation, the node’s representation captures the structural information within its $k$-hop neighborhood. The $k$-th layer of a GNN is shown as:
\begin{equation}
\begin{aligned}\label{equation-GNN-k-th-layer}
    a_v^{(k)} &= \mathsf{Aggregate}^{(k)}(\{ h_u^{(k-1)}: u \in \mathcal{N}(v) \})  \\
    h_v^{(k)} &= \mathsf{Combine}^{(k)} (h_v^{(k-1)}, a_v^{(k)})
\end{aligned}
\end{equation}
where $h_v^{(k)}$ is the feature vector of node $v$ at the $k$-th iteration, with $h_v^{(0)}$ initialized to $X_v$. The set $\mathcal{N}(v)$ denotes the neighbors of node $v$. The functions $\mathsf{Aggregate}^{(k)}(\cdot)$ and $\mathsf{Combine}^{(k)}(\cdot)$ are crucial components of GNN design.




GNNs have emerged as a powerful tool for POI recommendation~\cite{han2020stgcn,kim2021dynaposgnn,wang2021attentive,zhang2021leveraging,li2022attention,liu2024poi,zhang2024hyper}, effectively capturing complex relationships within location-based data. Earlier studies~\cite{han2020stgcn,kim2021dynaposgnn,wang2021attentive,zhang2021leveraging,li2022attention} focus on capturing spatial and temporal dynamics. Specifically, Han~\textit{et al.}~\cite{han2020stgcn} captures both spatial and temporal dynamics in user check-in data. It constructs a multigraph to represent contextual factors and their relationships, employs a time-aware sampling strategy to account for the evolving influence of neighboring nodes, and learns time-dependent node representations to reflect changing user preferences and POI popularity. Another study~\cite{kim2021dynaposgnn} introduces DynaPosGNN, a next POI recommendation model that integrates arrival times into user activities. Unlike traditional methods that rely solely on check-in history, DynaPosGNN analyzes the correlation between arrival time and two spatial dynamic graphs—the "User-POI graph" and the "POI-POI graph"—providing a more nuanced understanding of user behavior. ASGNN~\cite{wang2021attentive} incorporates a personalized hierarchical attention network to capture complex correlations between users and POIs in check-in sequences. This enables the model to capture both long-term and short-term user preferences, enhancing its ability to predict the next POI. Finally, Zhang \textit{et al.}~\cite{zhang2021leveraging} introduce GNN-POI, a POI recommendation framework that utilizes GNNs to learn node representations from both node information and topological structure, improving POI recommendation accuracy. GNNs' exceptional capacity to model complex relationships makes them well-suited for capturing the intricate connections between users and POIs in LBSNs.

\begin{figure}
\centering
\begin{tabular}{c c}
\hspace{-5.5mm}
  \begin{minipage}{0.245\textwidth}
	\includegraphics[width=\textwidth]{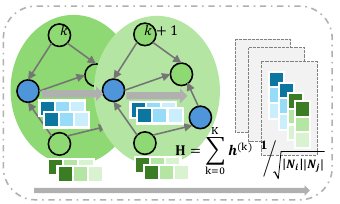}
  \end{minipage}\hspace{-3.0mm}
  &
  \begin{minipage}{0.20\textwidth}
    \includegraphics[width=\textwidth]{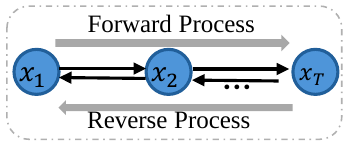}
  \end{minipage}\hspace{-3.0mm}
\\
(a) GNN model structure
&
(b) DM model sturcture
\end{tabular}
\caption{GNN and DM model structures.}
\label{fig:model_stru}
\vspace{-0.2in}
\end{figure}

Recent studies~\cite{zhang2024hyper,liu2024poi} based on GNN focus on using GNNs to capture side information. Liu \textit{et al.}~\cite{liu2024poi} propose a session-aware GNN to learn POI transfer preferences from similar users. This approach aims to understand how users' POI preferences evolve over time and how these preferences can be transferred to similar users. In contrast, HKGNN~\cite{zhang2024hyper} utilizes hypergraphs and side information to enhance POI recommendations. It employs a hypergraph to represent complex relationships between users, POIs, and other entities, a hypergraph neural network to learn from the structural information encoded in the hypergraph, and side information such as POI categories and user demographics to address data sparsity and improve recommendation accuracy.


\vspace{-0.15in}
\subsection{Self-Supervised Learning-Based}

Self-supervised learning (SSL)~\cite{munro1986self} is a versatile framework that leverages unsupervised data by creating surrogate tasks, known as pretext tasks, to help models learn meaningful representations. These tasks are designed to guide the learning process towards capturing useful features from the data. SSL methods are generally categorized into two approaches: generative and contrastive~\cite{liu2021self}. Generative SSL aims to create or reconstruct data, focusing on learning detailed representations, while contrastive SSL emphasizes comparing data samples by bringing similar pairs closer in the latent space and pushing dissimilar pairs apart.

The SSL process typically begins by defining a pretext task, which does not require labeled data but helps in learning valuable representations. In contrastive learning, the pretext task involves differentiating between augmented versions of the same image (positive pairs) and different images (negative pairs). The model is trained by minimizing the following loss function~\cite{chen2020simple}:
\begin{equation}
\begin{aligned}
    \mathcal{L}_{\text{SSL}} = -\log \frac{\mathsf{Exp}(\mathsf{Sim}(\mathsf{pair}_i, \mathsf{pair}_j)/\tau)}{\sum_{k=1}^{2N} \mathbf{1}_{[k \neq i]} \mathsf{Exp}(\mathsf{Sim}(\mathsf{pair}_i, \mathsf{pair}_k)/\tau)}
\end{aligned}
\end{equation}
where $\mathbf{1}_{[k \neq i]} \in \{0, 1\}$ is an indicator function that equals 1 only if $k \neq i$, $\mathsf{pair}_i$ and $\mathsf{pair}_j$ are the feature representations of a positive pair, $\mathsf{Sim}(\cdot, \cdot)$ denotes cosine similarity, $\tau$ is a temperature scaling parameter, and $N$ represents the number of data points.

SSL has emerged as a promising paradigm in POI recommendation~\cite{gui2024survey}. By leveraging patterns and structures within data, self-supervised techniques enable models to learn meaningful representations without explicit labels. In POI recommendation, these methods capture latent user preferences and spatial contexts from unlabeled data, enhancing personalized recommendations. Self-supervised POI recommendation systems~\cite{zhou2021self,li2022self,gao2022self,jiang2023modeling,wang2023exploring,wang2024graph,fu2024contrastive} can be divided into two categories. The first focuses on creating augmented views through human design~\cite{zhou2021self,li2022self,gao2022self}. Zhou \textit{et al.}~\cite{zhou2021self} introduced a Self-supervised Mobility Learning (SML) framework to enhance location-based tasks by encoding human mobility semantics. SML addresses challenges posed by sparse and noisy mobility data by leveraging spatio-temporal contexts and augmented traces. It generates contrastive views through methods such as sampling trajectory points. Another work~\cite{li2022self} introduces a paradigm that learns trajectory representations via random sampling of POI points, enhancing the model's understanding of mobility patterns. A Graph-enhanced Self-attentive layer uncovers transitional dependencies among POIs and captures temporal interests. Lastly, SelfTrip~\cite{gao2022self} focuses on trip recommendation using a two-step contrastive learning mechanism to learn nuanced representations of POIs and itineraries based on augmented views from random walks, further enhanced by four innovative trip augmentation methods.

Recent studies~\cite{jiang2023modeling,wang2023exploring,wang2024graph,fu2024contrastive} focus on automatically creating augmented views. While many studies prioritize POI recommendation performance, they often overlook the connection between POI sequences and contextual information, leading to limited data representations and less accurate recommendations. To address this, Jiang~\textit{et al.}~\cite{jiang2023modeling} propose LSPSL, a unified attention framework for next-POI recommendation that leverages self-supervised learning to model long- and short-term preferences. LSPSL uses a self-attention network and two self-supervised optimization objectives to explore relationships between POI sequences and contextual information, generating contrastive views automatically during pre-training. GSBPL~\cite{wang2023exploring} introduces graph-based data augmentation techniques for next-POI recommendation, modeling user behavior patterns via GNNs to create augmented trajectory graphs. By applying contrastive learning, GSBPL captures implicit behavioral patterns. To address limitations in capturing higher-order relationships and the varying importance of POIs, Fu~\textit{et al.}~\cite{fu2024contrastive} introduce SLS-REC, a self-supervised model for POI recommendation. SLS-REC leverages a spatio-temporal attention hypergraph neural network to capture spatial dependencies and the temporal evolution of short-term dynamic user interests while automatically creating contrastive views.

\vspace{-0.1in}
\subsection{Generative}
\vspace{-0.05in}
\subsubsection{Large Language Model-Based}

Large language models are advanced deep learning models with millions of parameters, excelling at understanding and generating human language with high accuracy and fluency. They perform tasks like translation, summarization, and question-answering by predicting word sequences. Depending on their architectural focus, LLMs can be broadly categorized into three types: encoder-only~\cite{devlin2019bert}, decoder-only~\cite{radford2019language}, and encoder-decoder models~\cite{raffel2020exploring}. Encoder-only LLMs focus on processing input text to create context-aware representations, excelling in tasks like classification, question answering, and information retrieval. Decoder-only LLMs specialize in text generation, predicting the next word in a sequence for tasks such as story writing and summarization. Encoder-decoder LLMs combine both components, processing input with an encoder and generating output with a decoder, making them ideal for tasks requiring both comprehension and generation.

Within the realm of POI recommendation, recent studies showcase the promising capabilities of LLMs. Feng \textit{et al.}~\cite{feng2024move} explore the use of LLMs, specifically ChatGPT, for predicting a user's next check-in location. Their work focuses on: (1) Prompt Engineering: Designing effective prompting strategies to elicit accurate location predictions from the LLM; (2) Ranking-Based Recommendation: Framing the recommendation task as a ranking problem, where the LLM ranks potential POIs based on their likelihood of being the user's next destination; and (3) Incorporating Spatio-Temporal Factors: Considering key factors influencing human mobility, such as user preferences, spatial distances, and sequential transitions, in the prediction process. Different from~\cite{feng2024move}, Li~\textit{et al.}~\cite{li2024large} investigate how pretrained LLMs can effectively leverage the rich contextual information embedded within LBSN data. This method aims to: (1) Preserve Data Heterogeneity: Maintaining the original format of LBSN data, thereby preventing the loss of valuable contextual information often encountered during preprocessing steps; and (2) Context-Aware Representation Learning: Utilizing the LLM's ability to capture complex relationships within unstructured data to learn more nuanced and context-aware representations of users and POIs.


\subsubsection{Diffusion Model-Based}

Diffusion models (DMs)~\cite{sohl2015deep} are a class of generative models that create data by gradually transforming a simple distribution into a complex one through a sequence of stochastic steps. Denoising Diffusion Probabilistic Models (DDPMs)~\cite{ho2020denoising}, a specific type of DM, refine this process by learning to reverse the diffusion, progressively removing noise from a noisy signal to generate high-quality samples. DDPMs consist of two main phases: the forward and the reverse diffusion process, shown in Fig.~\ref{fig:model_stru} (b).


DMs have emerged as a powerful tool for generating high-quality, diverse data across various domains. Recent research explores their application to POI recommendation~\cite{qin2023diffusion,wang2024dsdrec,zuo2024diff,long2024diffusion}, leveraging their ability to model complex distributions and capture intricate patterns in user behavior. Long \textit{et al.}~\cite{long2024diffusion} introduce DCPR, a collaborative learning framework utilizing diffusion models for the next POI recommendation. Key features include a Cloud-Edge-Device architecture, region-specific recommendations, and collaborative learning. Zuo \textit{et al.}~\cite{zuo2024diff} propose Diff-POI, a model leveraging diffusion models to capture users' spatial preferences for POI recommendation. Notable aspects include spatial preference modeling, custom graph encoding, and diffusion-based sampling techniques. Wang \textit{et al.}~\cite{wang2024dsdrec} present DSDRec, a recommender system combining diffusion models with semantic information extraction for next POI recommendation. Key highlights include the integration of semantic information extraction, pretrained language model enhancements, and spatio-temporal area encoding.


\subsubsection{Variational Autoencoder}

The variational autoencoder (VAE)~\cite{kingma2013auto,kingma2019introduction} is a unique type of autoencoder that introduces regularization to the encoding distribution during training, resulting in a well-structured latent space. VAEs offer a flexible framework for learning deep latent-variable models and their associated inference mechanisms.

Different VAEs~\cite{zhou2023uncertainty,li2022linking} offer unique advantages in POI recommendation, addressing various aspects of user behavior and data characteristics. WaPOIR~\cite{zhou2023uncertainty} emphasizes understanding the data distribution to enhance personalization. It integrates diverse data sources, including user preferences, social influences, and geographical information, employing a Wasserstein autoencoder to learn latent features of users and POIs, effectively capturing the overall data distribution. The model captures both long-term preferences from historical check-ins and short-term inclinations from recent visits. Li \textit{et al.}~\cite{li2022linking} employs a grid-based structure to incorporate spatial context, organizing check-in records and embedding users within their grid cells using a VAE. The decoder reconstructs check-in patterns, and a classifier associates representations with users, facilitating a structured approach to user modeling.

\subsection{Discussion of Models}

In this section, we compare the model types discussed earlier: Latent Factor Models (LFMs), Classic Neural Networks (NNs), Self-Supervised Learning (SSL), and Generative Models (Gen-Models), focusing on model effectiveness, inference efficiency, training scalability, and result explainability, as summarized in Table~\ref{tab:dis}.

\begin{table}[htbp]
    \setlength{\tabcolsep}{5pt}
    \centering
    \caption{Comparison of models based on effectiveness, efficiency, scalability, and explainability}
    \small
    \begin{tabular}{c|c|c|c|c}
    \toprule
    Criterion & LFMs & NNs & SSL & Gen-Models \\ \midrule
    Effectiveness  & Moderate  & High      & High     & High  \\
    Efficiency     & High      & Varies    & Varies   & Low   \\
    Scalability    & High      & Moderate  & Moderate & Low   \\
    Explainability & Moderate  & Low       & Moderate & High  \\ 
    \bottomrule
    \end{tabular}
    \label{tab:dis}
    \vspace{-0.1in}
\end{table}

Firstly, in terms of model effectiveness, LFMs provide moderate performance by using latent features to generate reasonable recommendations, though with limited depth. NNs, on the other hand, excel at capturing complex relationships, leading to more nuanced recommendations. SSL is particularly effective when labeled data is scarce, as it learns robust representations from unlabeled data. Gen-Models offer high effectiveness, especially when diverse recommendations are needed, due to their generative nature.

Secondly, regarding inference efficiency, LFMs are highly efficient, particularly with smaller datasets. NNs' efficiency depends on the network architecture, with larger models resulting in slower inference. The efficiency of SSL varies depending on the method used, with some being computationally expensive. Gen-Models tend to be less efficient due to their complexity and large parameter volumes, which result in longer inference times.

Additionally, in terms of training scalability, LFMs are highly scalable due to their simple structure, enabling efficient handling of large datasets. NNs and SSL have moderate scalability, as their training becomes computationally intensive with larger datasets. Gen-Models face significant scalability challenges, as training them becomes prohibitively expensive with the increasing data size.

Finally, in terms of result explainability, LFMs offer moderate transparency, providing some insight into user preferences through latent factors. NNs generally lack explainability due to their complex architectures, making it difficult to interpret their decisions. SSL provides moderate explainability, though interpreting the learned representations often requires further analysis. Gen-Models provide higher explainability than the other models, but understanding them fully necessitates exploring the latent space, which can be complex.

\section{Architecture}
\label{sec:archi}

In this section, we provide illustrations on \textbf{architectures}, including centralized-based, decentralized-based, and federated learning-based architectures. 

\vspace{-0.1in}
\subsection{Centralized Architecture}

Centralized architecture provides a scalable, cost-effective foundation for recommendation systems by using centralized data centers to manage computational resources. This model allows enterprises to respond quickly to demand changes, reduce overhead, and streamline maintenance. Centralizing data processing and storage enhances efficiency in managing data flow and system integration. The flexibility of centralized architecture is expressed by the following equation:

\begin{equation}
\begin{aligned}
    &\text{Centralized Efficiency} = \\
    &\frac{\text{Scalability} + \text{Cost Efficiency} + \text{Accessibility}}{\text{Initial Hardware Investments} + \text{Physical Constraints}}
\end{aligned}
\end{equation}

This equation illustrates how centralized systems optimize scalability, cost-effectiveness, and resource availability with minimal upfront hardware investments. We explore studies~\cite{wang2018exploiting,yin2017spatial,yang2022getnext,zhao2020go,wang2023adaptive} on POI recommendation systems developed using centralized architecture. For instance, Wang \textit{et al.}~\cite{wang2018exploiting} leveraged centralized computing to model geographical influences in POI recommendations with GeoIE. Centralized servers enable real-time processing of geographical data, enhancing the efficiency of recommendations by factoring in geo-influence, geo-susceptibility, and physical distance. Similarly, Yin \textit{et al.}~\cite{yin2017spatial} developed SH-CDL, where centralized architecture plays a crucial role in managing the fusion of diverse POI features and user preferences across multiple locations. The centralization of data allows SH-CDL to more effectively model both individual and public preferences by leveraging collective data from a large number of users. Furthermore, Wang \textit{et al.}~\cite{wang2023adaptive} introduced AGRAN, which benefits from centralized data integration to create adaptive POI graphs and fuse POI representations with dynamic spatial-temporal data, ensuring real-time responsiveness to evolving user preferences. These studies underscore the advantages of centralized architecture in POI recommendation systems, emphasizing enhanced efficiency and scalability. However, as the complexity of these systems increases, continuous innovation in defense mechanisms is essential to safeguard data integrity and user privacy within centralized frameworks.

\vspace{-0.1in}
\subsection{Decentralized Architecture}\label{subsec:decentralized}

The architecture of decentralized methodologies prioritizes the maximal utilization of local storage and processing capabilities, thereby conserving bandwidth and curtailing cloud storage expenses. Moreover, by processing data locally, these methodologies bolster data security and user privacy, mitigating the risks linked to data breaches and unauthorized access. Mathematically, the efficacy of decentralized methods can be encapsulated in the equation:

\vspace{-0.1in}
\begin{equation}
\begin{aligned}
    &\text{Decentralized Efficiency} = \\
    &\frac{\text{Local Processing Power} + \text{Storage Capacity}}{\text{Bandwidth Consumption} + \text{Cloud Storage Costs}}
\end{aligned}
\end{equation}
where the decentralized efficiency metric integrates local processing power, storage capacity, bandwidth consumption, and cloud storage expenses to gauge the overall effectiveness of decentralized computing solutions in comparison to cloud-based alternatives. 

In this part, we illustrate the decentralized architecture for POI recommendation. While centralized architectures excel in accuracy, they require substantial computational resources during training. To address these challenges, researchers have introduced decentralized architectures like~\cite{long2023model,long2023decentralized,zheng2024decentralized}. Specifically, Long \textit{et al.}~\cite{long2023model} propose Model-Agnostic Collaborative Learning (MAC), which allows users to tailor model architectures, such as adjusting dimensions and the number of hidden layers. To mitigate the scarcity of user data on devices, MAC~\cite{long2023model} pre-determines collaborators based on physical proximity, preferences, and social connections. It integrates insights from collaborators using a knowledge distillation approach, maximizing mutual information, and selectively sh-

\vspace{-1\baselineskip}  
\begin{center}
\scriptsize
\setlength{\tabcolsep}{0.3mm}{
\onecolumn
\begin{longtable}{c|c|c|c|c}
\captionsetup{justification=centering, width=\textwidth}
\caption{Related works in various application sub-tasks. We provide details on journal/conference names, generative techniques used, specific sub-tasks addressed, and evaluation datasets.}
\label{tab:gen_method}\\
    \toprule
    \textbf{Models} & \textbf{Source} & \textbf{Technique} & \textbf{Sub-task} & \textbf{Datasets} \\ \midrule
    \endfirsthead
    \hline
    \endfoot
    \rowcolor{lightgray}\multicolumn{5}{l}{\textbf{Latent Factor Model-based Models}} \\ \hline
    \multicolumn{5}{l}{\textbf{LDA-based Models}} \\ \hline
     LCLCARS~\cite{yin2013lcars} & \textit{KDD 2013}  & LDA & {\tkderevision{Location-Content-Aware Recommendation}} & Foursquare, and Douban Event\\ \hline
     LCARS~\cite{yin2014lcars} & \textit{TOIS 2014}  & LDA & Spatial Item Recommendation & Douban Event, and Foursquare \\ \hline
     JIM~\cite{yin2015joint} & \textit{CIKM 2015}  & LDA & POI Recommendation & Foursquare, and Twitter \\ \hline
     Geo-SAGE~\cite{wang2015geo} & \textit{KDD 2015}  & LDA & {\tkderevision{Spatial Item Recommendation}} & Foursquare, and Twitter\\ \hline
     TRM~\cite{yin2015joint} & \textit{CIKM 2015}  & LDA & POI Recommendation & Foursquare, and Twitter\\ \hline
     LA-LDA~\cite{yin2015modeling} & \textit{TKDD 2015}  & LDA & Location-based Recommendation & MovieLens, Gowalla, and Douban Event \\ \hline
     TRM+~\cite{yin2016joint} & \textit{TOIS 2016}  & LDA & POI Recommendation & Foursquare, and Twitter \\ \hline 
     SPORE~\cite{wang2016spore} & \textit{ICDE 2016}  & LDA & Spatial Item Recommendation & Foursquare, Twitter, and Synthetic Dataset \\ \hline 
     ST-LDA~\cite{yin2016adapting} & TKDE 2016 & LDA & POI Recommendation & Foursquare, and Gowalla\\ \hline 
     UCGT~\cite{yin2016discovering} & \textit{ICDE 2016}  & LDA & Social Community Detection & Foursquare, and Douban Event \\ \hline
     LSARS~\cite{wang2017location} & \textit{KDD 2017}  & LDA & Spatial Item Recommendation & Yelp, and Foursquare \\ \hline
     ST-SAGE~\cite{wang2017st} & \textit{TIST 2017}  & LDA & Spatial Item Recommendation & Foursquare, and Twitter \\ \hline
     TPM~\cite{wang2018tpm} & \textit{TIST 2018}  & LDA & Spatial Item Recommendation & Foursquare, Twitter, and Synthetic Dataset \\ \hline
    \multicolumn{5}{l}{\textbf{MF-based Models}} \\ \hline
     GeoMF~\cite{lian2014geomf} & \textit{KDD 2014}  & MF & POI Recommendation & Jiepang\\ \hline
     GeoMF++~\cite{lian2018geomf++} & \textit{TOIS 2018}  & MF & POI Recommendation & Gowalla, and Jiapang\\ \hline
     LGLMF~\cite{rahmani2020lglmf} & \textit{AIRS 2019}  & MF  & POI Recommendation & Foursquare, and Gowalla\\ \hline
     SSTPMF~\cite{davtalab2021poi} & \textit{KIS 2021}  & MF & POI Recommendation & Foursquare-TKY, and Gowalla-NYC\\ \hline
     SQPMF~\cite{wang2024sqpmf} & \textit{Applied Intelligence 2024}  & MF & POI Recommendation & GowallaFootnote, Foursquare, and Brightkite \\ \hline
    \rowcolor{lightgray}\multicolumn{5}{l}{\textbf{Classic Neural Network (NN)-based Models}} \\ \hline
    \multicolumn{5}{l}{\textbf{LSTM-based Models}} \\ \hline
    ST-LSTM~\cite{zhao2018go} & \textit{CoRR 2018}  & LSTM &POI Recommendation  &CA, Gowalla, SIN, Brightkite \\ \hline
    LSTM-S~\cite{zhan2019semantic} & \textit{MDM 2019}  & LSTM &POI Recommendation  &Foursquare \\ \hline
    PS-LSTM~\cite{zhong2021ps} & \textit{CNIT 2021}  & LSTM &POI Recommendation  &Yelp, Gowalla, Foursquare \\ \hline
    STMLA~\cite{zhang2022spatio} & \textit{ICWS 2022}  & LSTM &POI Recommendation  &Gowalla, Foursquare \\ \hline
    LSMA~\cite{yang2022attention} & \textit{IJGI 2022}  & LSTM & Next POI Recommendation & Foursquare-Charlotte (CHA), Foursquare-NYC \\ \hline
    AT-LSTM~\cite{lai2023poi} & \textit{IJSWIS 2023}  & LSTM &POI Recommendation  &Gowalla, Foursquare \\ \hline
    ATSD-GRU~\cite{jia2023attention} & \textit{IJITSA 2023}  & LSTM & POI Recommendation & mafengwo \\ \hline
    DLAN~\cite{wu2024dlan} & \textit{IFS 2024}  & LSTM &POI Recommendation  & Foursquare-NYC, Foursquare-TKY \\ \hline
    \multicolumn{5}{l}{\textbf{Transformer-based Methods}} \\ \hline
    TLR-M~\cite{halder2021transformer} & \textit{PAKDD 2021}  & Transformer & POI Recommendation  & Foursquare-NYC, Foursquare-TKY \\ \hline
    GETNext~\cite{yang2022getnext} & \textit{SIGIR 2022}  & Transformer & POI Recommendation  & Foursquare-NYC, Foursquare-TKY, CA \\ \hline
    CARAN~\cite{hossain2022caran} & \textit{IEEE Access 2022}  & Transformer & Next POI Recommendation & Foursquare-NYC, Foursquare-TKY, and Gowalla\\ \hline
    JANICP~\cite{zhong2022joint} & \textit{DSE 2022}  & Transformer & POI Recommendation & Weeplaces, Foursquare-NYC and Foursquare-TKY \\ \hline
    Li \textit{et al.}~\cite{li2022using} & \textit{IJGI 2022}  & Transformer & Next POI Recommendation & Foursquare-NYC and Foursquare-TKY \\ \hline
    STUIC-SAN~\cite{li2022spatio} & \textit{IJGI 2022}  & Transformer & Next POI Recommendation & Foursquare, Gowalla \\ \hline
    AMACF~\cite{zang2021cha} & \textit{IJGI 2022}  & Transformer & Next POI Recommendation & Foursquare‐NYC, Foursquare-TKY and Weeplaces \\ \hline
    CHA~\cite{wang2022next} & \textit{TOIS 2022}  & Transformer & Next POI Recommendation & Foursquare‐NYC, Foursquare-TKY \\ \hline
    HAT~\cite{wu2023reason} & {\tkderevision{\textit{TMM 2023}}}  & Transformer &POI Recommendation  & BJ, SH, NJ, CD \\ \hline 
    STAR-HiT~\cite{xie2023hierarchical} & \textit{ACM TIS 2023}  & Transformer &POI Recommendation  &Foursquare‐NYC, Foursquare‐US, Gowalla\\ \hline
    CAFPR~\cite{halder2023capacity} & \textit{Applied SC 2023}  & Transformer &POI Recommendation  & Foursquare-TKY, CA, Budapest, Melbourne, Magic k \\ \hline
    FPG~\cite{he2023feature} & \textit{Applied SC 2023}  & Transformer &POI Recommendation  & Foursquare, Gowalla \\ \hline
    TGAT~\cite{jiang2023temporal} & \textit{IF Systems 2023}  & Transformer &POI Recommendation  & Foursquare-NYC, Foursquare-TKY \\ \hline
    MobGT~\cite{xu2023revisiting} & \textit{SigSpatial 2023}  & Transformer &POI Recommendation  & Foursquare-NYC, Gowalla \\ \hline
    POIBERT~\cite{ho2022poibert} & \textit{IEEE Big data 2022}  & Transformer &POI Recommendation  &Budapest, Delhi, Edinburgh, Glasgow, Osaka, Perth, Toronto \\ \hline
    AutoMTN~\cite{qin2022next} & \textit{SIGIR 2022}  & Transformer &POI Recommendation  & Foursquare-NYC, Foursquare-TKY \\ \hline
    CCDSA~\cite{wang2023context} & \textit{Applied Intelligence 2023}  & Transformer & Next POI Recommendation & Foursquare-NY, Foursquare-TKY, Weeplaces-NY and Weeplaces-SF \\ \hline
    Liu \textit{et al.}~\cite{liu2023poi} & \textit{IJITSA 2023}  & Transformer & POI Recommendation & BrightKite \\ \hline
    TDGCN~\cite{cao2023improving} & \textit{IPM 2023}  & Transformer & Next POI Recommendation & Foursquare-TKY, Weeplaces and Gowalla-CA \\ \hline
    TGAT~\cite{jiang2023temporal} & \textit{JIFS 2023}  & Transformer & POI Recommendation & Foursquare-NYC, Foursquare-TKY \\ \hline
    MGAN~\cite{wu2023muti} & \textit{JIFS 2023}  & Transformer & POI Recommendation & Yelp, Foursquare \\ \hline
    STA-TCN~\cite{ou2023sta} & \textit{TKDE 2023}  & Transformer & Next POI Recommendation & Gowalla, Foursquare \\ \hline
    Xia \textit{et al.}~\cite{xia2023effective} & \textit{PerCom Workshops 2023}  & Transformer & Next POI Recommendation &Gowalla, Foursquare   \\ \hline
    STTF-Recommender~\cite{xu2023spatio} & \textit{IJGI 2023}  & Transformer &Next Location Recommendation  & Gowalla, Foursquare \\ \hline
    BayMAN~\cite{xia2023bayes} & \textit{TKDE 2023}  & Transformer & POI Recommendation  & Gowalla, NYC,  Foursquare \\ \hline
    Kumar \textit{et al.}~\cite{kumar2024modified} & \textit{IF 2024}  & Transformer &POI Recommendation  & Gowalla, Foursquare \\ \hline
    ImNext~\cite{he2024imnext} & \textit{KBS 2024}  & Transformer & Next POI Recommendation  &Gowalla, Foursquare \\ \hline
    BSA-ST-Rec~\cite{cheng2024point} & \textit{MTA 2024}  & Transformer & POI Recommendation  &Gowalla, Foursquare \\ \hline
    Wu \textit{et al.}~\cite{wu2024reason} & \textit{TMM 2024}  & Transformer & POI Recommendation  & Ctrip, Baidu Travel, Qunar, Yelp \\ \hline
    ROTAN~\cite{feng2024rotan} & \textit{KDD 2024}  & Transformer & Next POI Recommendation  & Foursquare-NYC, Foursquare-TKY, Gowalla-CA \\ \hline

    \multicolumn{5}{l}{\textbf{Graph Neural Network (GNN)-based Methods}} \\ \hline
    STGCN~\cite{han2020stgcn} & \textit{ICDM 2020}  & GCN & POI Recommendation  & Yelp, Boston, Chicago, London \\ \hline
    Dynaposgnn~\cite{kim2021dynaposgnn} & \textit{ICDM Workshops 2021}  & GNN &POI Recommendation  &Gowalla, Foursquare\\ \hline
    ASGNN~\cite{wang2021attentive} & \textit{World Wide Web 2021}  & GNN & Next POI Recommendation  & Gowalla, FourSquare, and Brightkite \\ \hline
    GNN-POI~\cite{zhang2021leveraging} & \textit{Neurocomputing 2021}  & GNN & POI Recommendation  & Gowalla, Foursquare, and Yelp \\ \hline
    ADQ-GNN~\cite{wang2021adq} & \textit{WISE 2021}  & GNN &POI Recommendation  & Foursquare-NYC, Foursquare-TKY, Gowalla \\ \hline
    GN-GCN~\cite{mo2022gn} & \textit{ICIIW 2022}  & GCN &POI Recommendation  & Gowalla, Yelp \\ \hline
    ATST-GGNN~\cite{li2022attention} & \textit{IEEE Access 2022}  & GNN & Next POI Recommendation & Foursquare, Gowalla \\ \hline
    Zhao \textit{et al.}~\cite{zhao2023poi} & \textit{MDM 2023}  & GNN & POI Recommendation  & Wi-fi login \\ \hline
    HS-GAT~\cite{zhang2024hybrid} & \textit{Expert Systems 2024}  & GNN & POI Recommendation  &Yelp, Boston, Chicago, London \\ \hline
    PROG-HGNN~\cite{meng2024poi} & \textit{Expert Systems 2024}  & GNN & POI Recommendation  & Foursquare, Gowallam, and Yelp \\ \hline
    CGNN-PRRG~\cite{liu2024poi} & \textit{IPM 2024}  & GNN & POI Recommendation  & Foursquare, Gowalla, and Yelp \\ \hline
    HKGNN~\cite{zhang2024hyper} & \textit{World Wide Web 2024}  & GNN & Next POI Recommendation & Foursquare-NYC, Foursquare-JK, Foursquare-KL, and Foursquare-SP \\ \hline
    \rowcolor{lightgray}\multicolumn{5}{l}{\textbf{Self Supervised Learning (SSL)-based Models}} \\ \hline
    SML~\cite{zhou2021self} & \textit{KBS 2021}  & SSL & Next Location Prediction & Foursquare, and Gowalla \\ \hline
    S2GRec~\cite{li2022self} & \textit{arXiv 2022}  & SSL & Next POI Recommendation & Gowalla, Foursquare-TKY and Foursquare-NYC \\ \hline
    SSTGL~\cite{liu2023self} & \textit{Applied Sciences 2023}  & SSL & POI Recommendation & Foursquare, Gowalla, and Meituan \\ \hline
    SelfTrip~\cite{gao2022self} & \textit{KBS 2022}  & SSL & Trip Recommendation & Toronto, Osaka, Glasgow, and Edinburgh \\ \hline
    GSBPL~\cite{wang2023exploring} & \textit{Electronics 2023}  & SSL & Next POI Recommendation & Gowalla, Foursquare-TKY, and Foursquare-NYC \\ \hline
    LSPSL~\cite{jiang2023modeling} & \textit{TKDD 2023}  & SSL &  Next POI Recommendation  & Foursquare-NYC and Foursquare-TKY \\ \hline
    Gao \textit{et al.}~\cite{gao2023predicting} & \textit{TKDE 2023}  & SSL &  Human Trajectory Prediction  & Foursquare-NYC, Foursquare-TKY, Los Angeles, Houston \\ \hline
    Wang \textit{et al.}~\cite{wang2024graph} & \textit{CSCWD 2024}  & SSL & POI Recommendation & Foursquare, and Gowalla \\ \hline
    SLS-REC~\cite{fu2024contrastive} & \textit{Expert Systems 2024}  & SSL & POI Recommendation & Foursquare, and Gowalla \\ \hline
    CLLP~\cite{zhou2024cllp} & \textit{SIGIR 2024}  & SSL & Next POI Recommendation & Foursquare, and Gowalla \\ \hline
    {\tkderevision SCL~\cite{chen2025self}}  & {\tkderevision \textit{Expert Systems 2025}}  & {\tkderevision SSL} & {\tkderevision POI Recommendation} & {\tkderevision Florence, Rome, Pisa, Edinburgh, and Toront} \\ \hline
    
    \rowcolor{lightgray}\multicolumn{5}{l}{\textbf{Generative-based Methods}} \\ \hline
    \multicolumn{5}{l}{\textbf{LLM-based Methods}} \\ \hline
    LLMmove~\cite{feng2024move} & \textit{CAI 2024}  & LLM & POI Recommendation & NYC, TKY \\ \hline
    Li \textit{et al.}~\cite{li2024large} & \textit{SIGIR 2024}  & LLM & POI Recommendation & NYC, TKY, Gowalla-CA \\ \hline
    \multicolumn{5}{l}{\textbf{Diffusion Models (DMs)-based Methods}} \\ \hline
    Diff-POI~\cite{qin2023diffusion} & \textit{TOIS 2023}  & Diffusion & Next POI Recommendation & Gowalla, Foursquare-SIN, Foursquare-TKY, and Foursquare-NYC \\ \hline
    DSDRec~\cite{wang2024dsdrec} & \textit{Information Sciences 2024}  & Diffusion & Next POI Recommendation & Foursquare-NYC, and Foursquare-TKY \\ \hline
    Diff-DGMN~\cite{zuo2024diff} & \textit{IOT 2024}  & Diffusion & POI Recommendation & IST, JK, SP, NYC, LA \\ \hline
    DCPR~\cite{long2024diffusion} & {\tkderevision \textit{KDD 2024}}  & Diffusion & POI Recommendation & Foursquare, and Weeplace \\ \hline
    \multicolumn{5}{l}{\textbf{VAE-based Methods}} \\ \hline
    WaPOIR~\cite{zhou2023uncertainty} & \textit{TSMCS 2023}  & VAE & POI Recommendation & Gowalla, Foursquare \\ \hline
    Li \textit{et al.}~\cite{li2022linking} & \textit{ICAI 2022}  & VAE & POI Recommendation & Gowalla, Foursquare \\ \hline
\end{longtable}}
\end{center}
\begin{multicols}{2}

\noindent aring informed decisions on a reference dataset rather than sensitive models or gradients. However, current collaborative learning systems rely on a single shared reference dataset, which may not align with individual preferences and hinder knowledge exchange. To address this, Zheng \textit{et al.}~\cite{zheng2024decentralized} propose the Adaptive Reference Data for Decentralized Collaborative Learning (DARD), which creates adaptive reference datasets for improved collaboration. DARD uses transformation and probability data generation techniques to create a public reference pool, from which personalized reference data is selected dynamically for each user based on loss tracking and influence function analysis. Users train models with private data and collaborate with spatial and contextual neighbors, exchanging soft decisions based on their personalized reference data.

{\tkderevision{While both MAC and DARD have merits, they suit different scenarios. MAC is effective when a reference dataset is available, such as in cases of public check-in sequences shared under authorization. DARD is preferred when no reference data exists, as it creates adaptive datasets for individual preferences. Each method excels under specific conditions, and neither method universally outperforms the other.}}

\vspace{-0.1in}
\subsection{Federated Learning-based Architecture}\label{subsec:fl-based}

Federated Learning (FL)~\cite{mcmahan2017communication} revolutionizes machine learning model training by enabling a decentralized and collaborative approach that ensures data privacy. This innovative technique allows models to be trained locally on devices without sharing raw data, thus providing a secure solution for deploying new machine learning applications. Fig.~\ref{fig:central-decentral-FL} presents a comparison between the centralized, decentralized, and FL-based architectures.

The FL-based approach offers the advantage of reducing data transfer and mitigating privacy risks by keeping sensitive information on local devices. The efficacy of Federated Learning can be mathematically captured through the following equation:
\begin{equation}
\begin{aligned}
    &\text{FL Efficiency} = \\
    &\frac{\text{DPP} + \text{Computational Resource Optimization}}{\text{Bandwidth Utilization} + \text{Model Update Aggregation}}
\end{aligned}
\end{equation}
Where $\text{DPP}$ denotes Data Privacy Protection.
This formula encapsulates how Federated Learning optimizes data privacy protection, computational resources, bandwidth usage, and model update aggregation to bolster the efficiency of machine learning model training across distributed devices.

\begin{figure}[H]
    \centering
    \setlength{\belowcaptionskip}{-0.25cm}
    \includegraphics[width=0.45\textwidth]{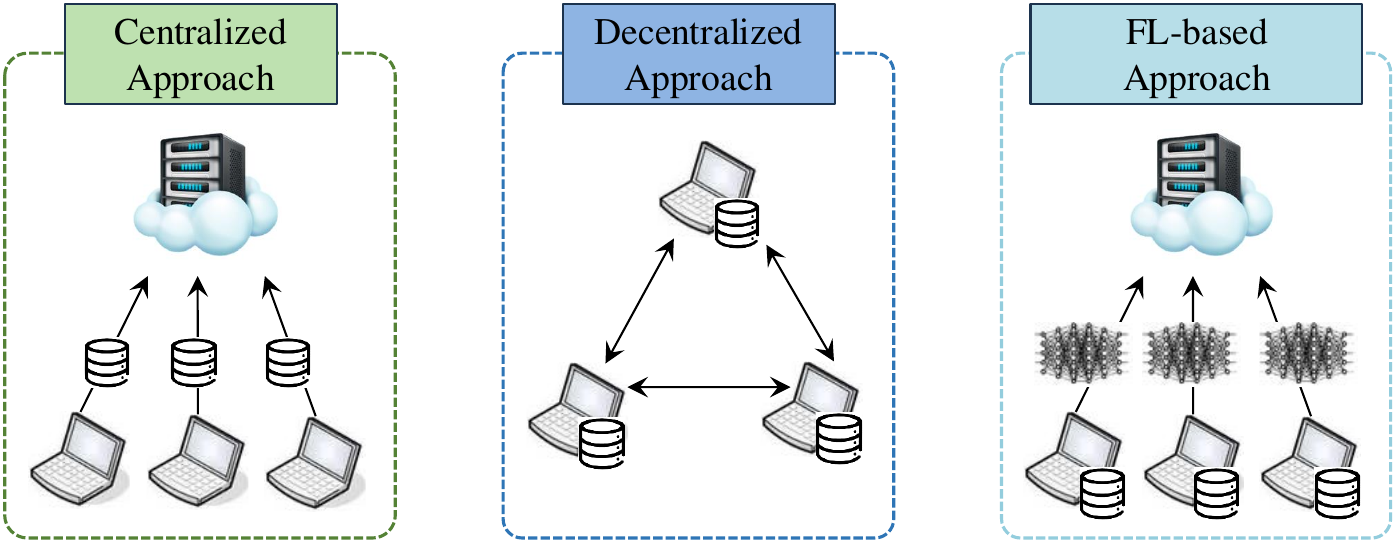}
    \caption{The centralized, decentralized, and FL-based approaches.}
    \label{fig:central-decentral-FL}
\end{figure}

FL operates on the principle of training machine learning models without requiring access to individual users' raw private data~\cite{wang2021poi,huang2022geographical,ye2023adaptive,zhong2024scfl,dong2024sfl,perifanis2023fedpoirec,zhang2023fine,guo2021prefer,dong2023sequential,an2024nrdl}. An earlier work~\cite{wang2021poi} introduces a cross-domain POI framework that incorporates federated learning and privacy safeguards. It leverages data from related domains to address the cold-start issue and employs federated learning to analyze users' historical data locally, encrypting feature distributions to ensure privacy while transferring knowledge. This method addresses challenges like sensitive POI data and cold-start problems caused by data scarcity. Another work presents PREFER~\cite{guo2021prefer}, a federated learning framework for POI recommendation with edge acceleration. Users develop individualized models locally and exchange multi-dimensional, user-agnostic parameters for privacy. These parameters are then consolidated on edge servers, enhancing recommendation speed. Huang \textit{et al.}~\cite{huang2022geographical} propose a federated learning-based geographical POI recommendation approach, framing the task as a matrix factorization problem solved through singular value decomposition and stochastic gradient descent, with only gradients transmitted for privacy protection. In addition, a privacy-focused framework, named FedPOIRec~\cite{perifanis2023fedpoirec}, was proposed to integrate the preferences of a user's friends after the federated computation. In this framework, data remains on the user's device, and updates are aggregated by a parameter server, and it allows users to exchange learned parameters with friends to improve personalization by using a CKKS fully homomorphic encryption scheme to ensure privacy during this process.

{\tkderevision{Each method has its strengths depending on the scenario. The cross-domain POI framework~\cite{wang2021poi} is useful when related domain data is available to address cold-start issues. PREFER~\cite{guo2021prefer} excels in environments requiring privacy and fast, localized recommendations using edge servers. Huang \textit{et al.}'s~\cite{huang2022geographical} method is ideal when geographical data is crucial, while FedPOIRec~\cite{perifanis2023fedpoirec} is best for applications relying on social connections for personalized recommendations. Each approach offers advantages under specific conditions.}}

\section{Security}
\label{sec:security}

In this section, we provide illustrations on \textbf{security}, including data integrity threats and user privacy protection in POI recommender systems.

\subsection{{\tkderevision{System Vulnerability Analysis}}}
Data integrity threats refer to malicious activities that compromise the accuracy, consistency, and trustworthiness of data within systems, particularly those that rely heavily on data-driven technologies. These threats can manifest at various stages, including data storage, transmission, and processing, significantly impacting the utility of POI recommender systems. 

One prominent data integrity threat in POI recommendation systems is poisoning attacks~\cite{rong2022fedrecattack,zhang2022loki}. These attacks manipulate training data, undermining the credibility and effectiveness of machine learning systems by injecting malicious data. Such attacks skew model predictions to benefit the attacker. Figure~\ref{fig:poisson-attack} illustrates the mechanisms of data and model poisoning, highlighting vulnerabilities from compromised datasets. The influence of poisoning attacks can be formally represented by manipulating training data, altering the learning process of machine learning models, as shown in the equation $\mathbf{X}_{\mathsf{poisoned}} = \mathbf{X}_{\mathsf{clean}} + \mathbf{A}_{\mathsf{perturb}}$, 
where $\mathbf{X}_{\mathsf{poisoned}}$ denotes the poisoned data, $\mathbf{X}_{\mathsf{clean}}$ represents the clean data, and $\mathbf{A}_{\mathsf{perturb}}$ signifies the adversarial perturbation introduced during the attack.

\begin{figure}[H]
    \vspace{-0.1in}
    \centering
    \setlength{\belowcaptionskip}{-0.25cm}
    \includegraphics[width=0.48\textwidth]{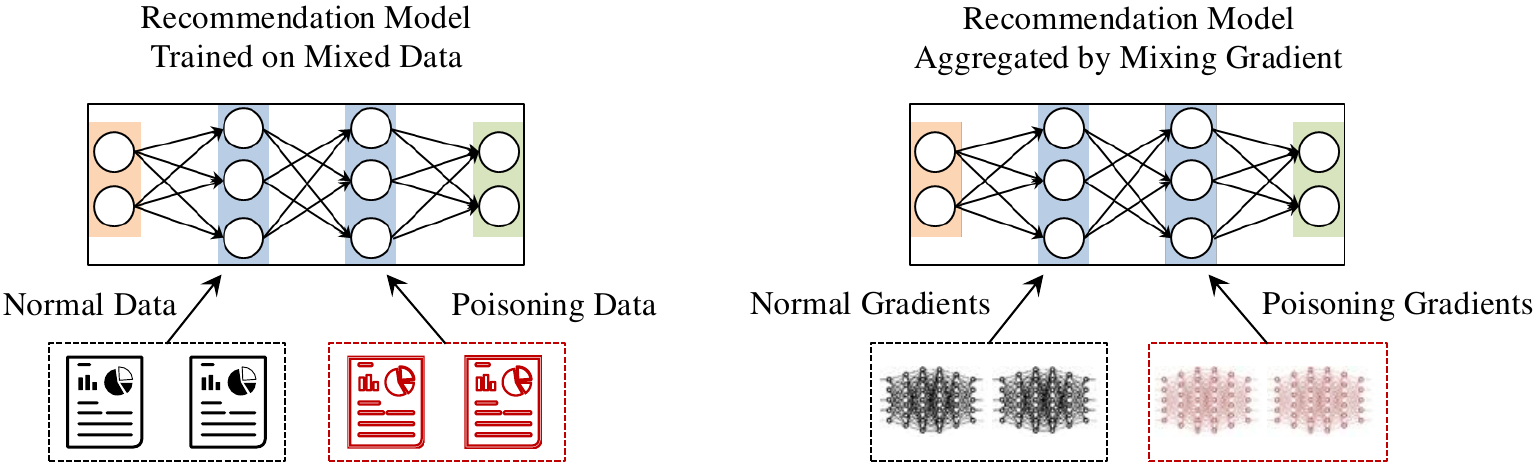}
    \caption{Data poisoning (Left) and model poisoning (Right)}
    \label{fig:poisson-attack}
\end{figure}



Rong~\textit{et al.}~\cite{rong2022fedrecattack} introduced FedRecAttack, a model poisoning attack targeting federated recommendation systems. By exploiting public interactions to approximate user feature vectors, attackers upload poisoned gradients through malicious users. Experiments show FedRecAttack is highly effective, even with minimal malicious users and public interactions, revealing FR's unexpected vulnerability. Zhang~\textit{et al.}~\cite{zhang2022loki} presented LOKI, a data poisoning attack on next-item POI recommendation systems. Using reinforcement learning, LOKI generates adversarial user behaviors, reducing retraining costs by interacting with a simulator. Results show LOKI surpasses previous attacks across diverse POI datasets and models. Another study~\textit{et al.}~\cite{yuan2023manipulating} introduced a new poisoning attack method that manipulates the ranking and exposure of top-K recommended targets in recommendation systems. Unlike previous methods, this attack operates without prior system knowledge, employing synthetic malicious users who strategically upload poisoned gradients targeting products related to desired items.

Additionally, POI recommendation systems may face various other threats, including Sybil attacks and the physical trajectory inference attack. Wang~\textit{et al.}~\cite{wang2018ghost} revealed vulnerabilities in real-time crowdsourced maps, demonstrating how software-based Sybil devices can exploit weak location authentication. Their techniques enable large-scale generation of these devices, leading to significant security and privacy threats in systems like Waze. Recently, Long~\textit{et al.}~\cite{long2024physical} introduced a novel physical trajectory inference attack (PTIA) that exploits aggregated information from POIs to reveal users' historical trajectories in decentralized recommendation.

These studies highlight the increasing sophistication of attacks on POI recommendation systems, revealing critical vulnerabilities. As attackers refine their methods, robust and adaptive defenses must evolve to protect the integrity and trustworthiness of POI recommendation processes.

\vspace{-0.1in}
\subsection{User Privacy Protection}

In recent years, privacy concerns have become critical issues in POI recommendation systems. These systems systems process vast amounts of sensitive user information to generate personalized recommendations, making them attractive targets for malicious actors seeking to exploit this data.


To address privacy concerns, various cryptographic techniques and privacy-preserving mechanisms are essential for safeguarding sensitive information in POI recommendation systems. Decentralized secure computation~\cite{chen2018privacy,chen2020practical} enables multiple parties to collaboratively compute a function over their inputs while keeping them private. Chen \textit{et al.}~\cite{chen2018privacy} proposed a decentralized matrix factorization framework for POI recommendation, addressing privacy and computational issues by enabling decentralized training on user devices. This approach preserves users' rating data and enhances recommendation accuracy. Furthermore, Chen \textit{et al.}~\cite{chen2020practical} introduced the PriRec framework, which maintains users' private data on their devices while employing local differential privacy techniques and secure decentralized gradient descent to protect model privacy, achieving comparable recommendation accuracy.

Additionally, homomorphic encryption~\cite{liu2017privacy,perifanis2023fedpoirec} is also commonly used for privacy-preserving POI recommendation. It allows computations on encrypted data without decryption, ensuring sensitive information remains secure. Figure~\ref{fig:HE} illustrates the workflow of homomorphic encryption, which conceals the underlying plaintext during processing. Li \textit{et al.}~\cite{liu2017privacy} developed a privacy-preserving framework for POI recommendation by utilizing homomorphic encryption. They address the cold start problem prevalent in POI recommendation systems, which arises due to the sparsity of user-location check-in data. Their framework leverages partially homomorphic encryption to enable trust-oriented recommendations while ensuring that sensitive user check-in and trust data remain private. By employing offline encryption and parallel computing, Li \textit{et al.}~\cite{liu2017privacy} ensured that their protocols efficiently protect the private data of all parties involved in the recommendation process. They also prove that their proposed protocols are secure against semi-honest adversaries, and their experiments on both synthetic and real datasets demonstrate the feasibility of privacy-preserving recommendations with acceptable computation and communication costs.

\begin{figure}[H]
    \centering
    \setlength{\belowcaptionskip}{-0.25cm}
    \includegraphics[width=0.49\textwidth]{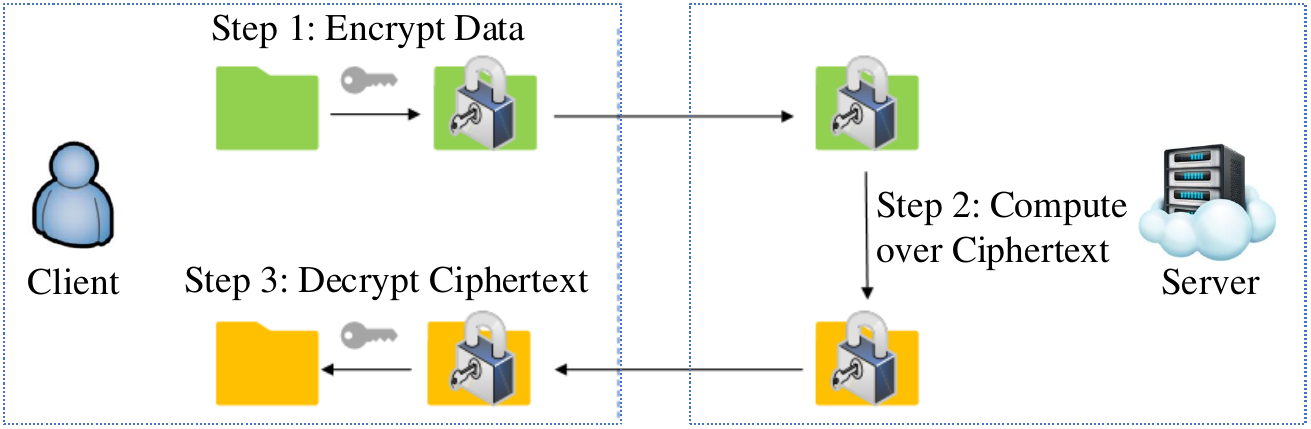}
    \caption{The workflow of homomorphic encryption}
    \label{fig:HE}
\end{figure}

\vspace{-0.1in}
\section{Future Directions}
\label{sec:future}

As POI recommendation systems continue to evolve, vast opportunities for innovation will keep emerging across the three pivotal dimensions: models, architectures, and security. By integrating cutting-edge model structures, adopting flexible architectures, and reinforcing security measures, the future of POI recommendation systems is expected to be more adaptive, personalized, and resilient to emerging challenges. Below, we outline the key future directions that will shape the development of these systems across these essential aspects.

\subsection{Models}
Future POI recommendation models will move beyond traditional and deep learning techniques, incorporating various data modalities and more advanced methodologies to further improve accuracy and personalization.
\textbf{(1)} \textbf{LLM Agent-Powered POI Recommendation.} The integration of Large Language Model (LLM) agents~\cite{xi2023rise} into POI recommendation systems offers another approach to enhance personalization POI recommendation. LLM agents can interpret user preferences through natural language interactions, making real-time POI recommendations based on conversational queries and feedback \cite{liu2024semantic,ning2024cheatagent}. These agents can provide personalized POI suggestions by engaging users in dialogue to clarify needs and generate tailored recommendations. By incorporating contextual factors such as location, time, and user-specific constraints, LLM-powered POI recommendation systems offer dynamic, context-aware suggestions. \textbf{(2) LLMs-Powered Explainable POI Recommendation.} Creating an explainable POI recommendation system via large language models is a promising avenue for leveraging advanced language processing capabilities to provide transparent and understandable POI recommendations. By utilizing these models, users can receive recommendations accompanied by explanations that clarify why a particular POI is suggested. This transparency enhances user trust and engagement by shedding light on the decision-making process behind each recommendation, ultimately improving the overall user experience of POI recommendation. {\tkderevision{\textbf{(3) Semantic Alignment Between POI Models and Real Applications.} Bridging semantic gaps between spatial data and real-world contexts enables POI models to understand POI spatial information. Integrating geographic hierarchies, proximity patterns, and spatial correlations among POIs (\eg, analyzing a park’s relation to transit hubs and neighborhoods) will improve POI recommendation contextual relevance, thereby improving POI recommendation accuracy.}}


\subsection{Architectures}
{\tkderevision{ Future POI recommendation architectures will primarily be distributed, incorporating technologies such as decentralized and federated recommendation. In actual deployment, these approaches face efficiency bottlenecks, making it difficult to effectively meet the demands of large-scale systems. The following outlines several future research directions focusing on scalability, latency, and communication efficiency.
\textbf{(1) Scalability Enhancement in Distributed POI Recommendation.} Improving the scalability of POI recommendation architectures is crucial for their effective deployment in large-scale systems. Current distributed systems struggle to accommodate an increasing number of computational nodes due to the overhead of large-scale data processing and inter-node coordination. The primary challenge lies in the need for efficient model training and result aggregation while maintaining computational efficiency. Future research could explore highly collaborative recommendation frameworks that dynamically balance computational workloads, optimize resource allocation, and simplify coordination among nodes to improve scalability and ensure efficient system expansion.
\textbf{(2) Latency Optimization for Real-Time POI Recommendation.} Reducing latency is a fundamental requirement for achieving real-time POI recommendations, particularly in decentralized collaborative systems. These architectures rely on multiple distributed nodes to perform recommendation tasks, but frequent data synchronization, inter-node communication, and model parameter exchanges introduce significant delays, impacting responsiveness. To mitigate these issues, future research could focus on developing asynchronous communication protocols, parallel computation strategies, and lightweight update mechanisms to enhance real-time performance in decentralized POI recommendation systems.
\textbf{(3) Communication-Efficient Federated POI Recommendation.} Federated recommendation models face significant challenges in handling the transmission of large-scale model parameters across distributed nodes~\cite{ye2023adaptive,dong2024sfl}. As model sizes increase, resource-constrained nodes may struggle to efficiently participate in global model training, limiting the scalability of federated learning. To address this challenge, future research could focus on designing communication-efficient aggregation algorithms, adaptive parameter compression techniques, and incentive-driven task distribution mechanisms to optimize the efficiency of federated POI recommendation and enable seamless collaboration across diverse computational environments.

}}

\end{multicols}

\clearpage
\begin{multicols}{2}

\subsection{Security}
{\tkderevision{ Future research needs to address the challenge of balancing security and efficiency in existing privacy-preserving POI recommendation systems, as traditional techniques such as federated learning, differential privacy, and cryptographic methods often encounter efficiency issues due to their high computational or communication costs. The following research directions are worth exploring.
\textbf{(1) Application-Oriented Privacy-Preserving POI Recommendation.} Optimizing privacy-preserving POI recommendation systems for specific application scenarios is crucial for improving efficiency. Different privacy protection methods exhibit varying trade-offs between security and computational cost. Federated learning excels in distributed training environments by safeguarding user privacy while minimizing data transmission overhead. Cryptographic techniques, such as homomorphic encryption and secure multi-party computation, are better suited for small-scale inference tasks that require strong privacy guarantees but lower computational complexity. Additionally, emerging machine unlearning techniques~\cite{li2024survey} offer efficient solutions for large-scale training and inference tasks, making them promising candidates for privacy-preserving POI recommendation. Future research should focus on designing adaptive frameworks that dynamically select the most suitable privacy-preserving strategy based on task characteristics such as training mode, inference complexity, and data scale. \textbf{(2) TEE-Based Privacy-Preserving POI Recommendation.} Leveraging Trusted Execution Environments (TEEs) presents a promising direction for enhancing security while maintaining computational efficiency in POI recommendation. TEEs, such as Intel SGX~\cite{costan2016intel} and Arm TrustZone~\cite{ngabonziza2016trustzone}, provide secure enclaves for executing privacy-sensitive computations with minimal performance overhead. However, they remain susceptible to side-channel attacks and hardware vulnerabilities. Future research should explore techniques to enhance TEE security, such as mitigating known attack vectors and integrating lightweight cryptographic enhancements. By improving the robustness of TEEs, privacy-preserving POI recommendation systems can achieve secure and efficient computation without significant trade-offs in performance. Additionally, emerging cross-domain technologies, such as data watermarking, could be integrated to further enhance both security and efficiency in privacy-preserving POI recommendation systems. By combining technological customization with hardware optimization, future systems can improve privacy protection effectiveness while maintaining efficiency.

}}

\section{Conclusion}
\label{sec:conclu}

The abundance of spatio-temporal data generated by smartphones and location-based social networks has fueled advancements in POI recommendation systems. These systems enhance user experiences, personalize interactions, and streamline decision-making in the digital realm. Our survey bridges the gap in existing literature by providing a comprehensive overview of recent advancements in POI recommendation, encompassing models, architectures, and security aspects. We highlight the emergence of advanced machine learning models, explore evolving system architectures, and emphasize the crucial need for robust security measures to protect user privacy and prevent malicious manipulation. This comprehensive taxonomy serves as a valuable resource for researchers and practitioners seeking to advance the field of POI recommendation and harness the full potential of spatio-temporal data.
\bibliographystyle{IEEEtran}
\bibliography{ref}

\begin{thebibliography}{100}
\providecommand{\url}[1]{#1}
\csname url@samestyle\endcsname
\providecommand{\newblock}{\relax}
\providecommand{\bibinfo}[2]{#2}
\providecommand{\BIBentrySTDinterwordspacing}{\spaceskip=0pt\relax}
\providecommand{\BIBentryALTinterwordstretchfactor}{4}
\providecommand{\BIBentryALTinterwordspacing}{\spaceskip=\fontdimen2\font plus
\BIBentryALTinterwordstretchfactor\fontdimen3\font minus \fontdimen4\font\relax}
\providecommand{\BIBforeignlanguage}[2]{{%
\expandafter\ifx\csname l@#1\endcsname\relax
\typeout{** WARNING: IEEEtran.bst: No hyphenation pattern has been}%
\typeout{** loaded for the language `#1'. Using the pattern for}%
\typeout{** the default language instead.}%
\else
\language=\csname l@#1\endcsname
\fi
#2}}
\providecommand{\BIBdecl}{\relax}
\BIBdecl

\bibitem{ye2010location}
M.~Ye, P.~Yin, and W.-C. Lee, ``Location recommendation for location-based social networks,'' in \emph{ACM SIGSPATIAL GIS}, 2010, pp. 458--461.

\bibitem{bao2015recommendations}
J.~Bao, Y.~Zheng, D.~Wilkie, and M.~Mokbel, ``Recommendations in location-based social networks: a survey,'' \emph{GeoInformatica}, vol.~19, pp. 525--565, 2015.

\bibitem{chorley2015personality}
M.~J. Chorley, R.~M. Whitaker, and S.~M. Allen, ``Personality and location-based social networks,'' \emph{Computers in Human Behavior}, vol.~46, pp. 45--56, 2015.

\bibitem{andrienko2007multimodal}
G.~Andrienko and N.~Andrienko, ``Multimodal analytical visualisation of spatio-temporal data,'' in \emph{Multimedia Cartography}.\hskip 1em plus 0.5em minus 0.4em\relax Springer, 2007, pp. 327--346.

\bibitem{sharafi2022novel}
M.~Sharafi, M.~Yazdchi, R.~Rasti, and F.~Nasimi, ``A novel spatio-temporal convolutional neural framework for multimodal emotion recognition,'' \emph{Biomedical Signal Processing and Control}, vol.~78, p. 103970, 2022.

\bibitem{du2023cross}
J.~Du, S.~Zhou, J.~Yu, P.~Han, and S.~Shang, ``Cross-task multimodal reinforcement for long tail next poi recommendation,'' \emph{IEEE TMM}, 2023.

\bibitem{zhang2024survey}
Q.~Zhang, H.~Wang, C.~Long, L.~Su, X.~He, J.~Chang, T.~Wu, H.~Yin, S.-M. Yiu, Q.~Tian \emph{et~al.}, ``A survey of generative techniques for spatial-temporal data mining,'' \emph{arXiv preprint arXiv:2405.09592}, 2024.

\bibitem{zhang2023online}
Q.~Zhang, Z.~Wang, C.~Long, C.~Huang, S.-M. Yiu, Y.~Liu, G.~Cong, and J.~Shi, ``Online anomalous subtrajectory detection on road networks with deep reinforcement learning,'' in \emph{IEEE ICDE}.\hskip 1em plus 0.5em minus 0.4em\relax IEEE, 2023, pp. 246--258.

\bibitem{zhang2024billiards}
Q.~Zhang, Z.~Wang, C.~Long, and S.-M. Yiu, ``Billiards sports analytics: Datasets and tasks,'' \emph{arXiv preprint arXiv:2407.19686}, 2024.

\bibitem{zhang2025efficient}
Q.~Zhang, X.~Gao, H.~Wang, S.-M. Yiu, and H.~Yin, ``Efficient traffic prediction through spatio-temporal distillation,'' \emph{arXiv preprint arXiv:2501.10459}, 2025.

\bibitem{zhang2024graph}
Q.~Zhang, L.~Xia, X.~Cai, S.-M. Yiu, C.~Huang, and C.~S. Jensen, ``Graph augmentation for recommendation,'' in \emph{IEEE ICDE}.\hskip 1em plus 0.5em minus 0.4em\relax IEEE, 2024, pp. 557--569.

\bibitem{xu2024mmpoi}
Y.~Xu, G.~Cong, L.~Zhu, and L.~Cui, ``Mmpoi: A multi-modal content-aware framework for poi recommendations,'' in \emph{WWW}, 2024, pp. 3454--3463.

\bibitem{rahmani2022systematic}
H.~A. Rahmani, M.~Aliannejadi, M.~Baratchi, and F.~Crestani, ``A systematic analysis on the impact of contextual information on point-of-interest recommendation,'' \emph{TIS (TOIS)}, vol.~40, no.~4, pp. 1--35, 2022.

\bibitem{qian2019spatiotemporal}
T.~Qian, B.~Liu, Q.~V.~H. Nguyen, and H.~Yin, ``Spatiotemporal representation learning for translation-based poi recommendation,'' \emph{TIS (TOIS)}, vol.~37, no.~2, pp. 1--24, 2019.

\bibitem{blei2003latent}
D.~M. Blei, A.~Y. Ng, and M.~I. Jordan, ``Latent dirichlet allocation,'' \emph{Journal of machine Learning research}, vol.~3, no. Jan, pp. 993--1022, 2003.

\bibitem{mnih2007probabilistic}
A.~Mnih and R.~R. Salakhutdinov, ``Probabilistic matrix factorization,'' \emph{NIPS}, vol.~20, pp. 1257--1264, 2007.

\bibitem{yin2014lcars}
H.~Yin, B.~Cui, Y.~Sun, Z.~Hu, and L.~Chen, ``Lcars: A spatial item recommender system,'' \emph{TIS (TOIS)}, vol.~32, no.~3, pp. 1--37, 2014.

\bibitem{yin2015dynamic}
H.~Yin, B.~Cui, L.~Chen, Z.~Hu, and X.~Zhou, ``Dynamic user modeling in social media systems,'' \emph{TOIS}, 2015.

\bibitem{hochreiter1997long}
S.~Hochreiter and J.~Schmidhuber, ``Long short-term memory,'' \emph{Neural computation}, vol.~9, no.~8, pp. 1735--1780, 1997.

\bibitem{vaswani2017attention}
A.~Vaswani, N.~Shazeer, N.~Parmar, J.~Uszkoreit, L.~Jones, A.~N. Gomez, {\L}.~Kaiser, and I.~Polosukhin, ``Attention is all you need,'' in \emph{NIPS}, 2017, pp. 5998--6008.

\bibitem{kim2021dynaposgnn}
J.~Kim, S.~Jeong, G.~Park, K.~Cha, I.~Suh, and B.~Oh, ``Dynaposgnn: Dynamic-positional gnn for next poi recommendation,'' in \emph{ICDMW}.\hskip 1em plus 0.5em minus 0.4em\relax IEEE, 2021, pp. 36--44.

\bibitem{wang2021attentive}
D.~Wang, X.~Wang, Z.~Xiang, D.~Yu, S.~Deng, and G.~Xu, ``Attentive sequential model based on graph neural network for next poi recommendation,'' \emph{World Wide Web}, vol.~24, no.~6, pp. 2161--2184, 2021.

\bibitem{wang2021adq}
Y.~Wang, A.~Liu, J.~Fang, J.~Qu, and L.~Zhao, ``Adq-gnn: Next poi recommendation by fusing gnn and area division with quadtree,'' in \emph{Web Information Systems Engineering--WISE 2021}.\hskip 1em plus 0.5em minus 0.4em\relax Springer, 2021, pp. 177--192.

\bibitem{chang2024survey}
Y.~Chang, X.~Wang, J.~Wang, Y.~Wu, L.~Yang, K.~Zhu, H.~Chen, X.~Yi, C.~Wang, Y.~Wang \emph{et~al.}, ``A survey on evaluation of large language models,'' \emph{ACM TIST}, pp. 1--45, 2024.

\bibitem{yang2023diffusion}
L.~Yang, Z.~Zhang, Y.~Song, S.~Hong, R.~Xu, Y.~Zhao, W.~Zhang, B.~Cui, and M.-H. Yang, ``Diffusion models: A comprehensive survey of methods and applications,'' \emph{ACS}, vol.~56, no.~4, pp. 1--39, 2023.

\bibitem{ho2020denoising}
J.~Ho, A.~Jain, and P.~Abbeel, ``Denoising diffusion probabilistic models,'' \emph{NIPS}, pp. 6840--6851, 2020.

\bibitem{jaiswal2020survey}
A.~Jaiswal, A.~R. Babu, M.~Z. Zadeh, D.~Banerjee, and F.~Makedon, ``A survey on contrastive self-supervised learning,'' \emph{Technologies}, vol.~9, no.~1, p.~2, 2020.

\bibitem{long2023decentralized}
J.~Long, T.~Chen, Q.~V.~H. Nguyen, and H.~Yin, ``Decentralized collaborative learning framework for next poi recommendation,'' \emph{TIS}, vol.~41, no.~3, pp. 1--25, 2023.

\bibitem{meng2024poi}
L.~Meng, Z.~Liu, D.~Chu, Q.~Z. Sheng, J.~Yu, and X.~Song, ``Poi recommendation for occasional groups based on hybrid graph neural networks,'' \emph{Expert Systems with Applications}, vol. 237, p. 121583, 2024.

\bibitem{perifanis2023fedpoirec}
V.~Perifanis, G.~Drosatos, G.~Stamatelatos, and P.~S. Efraimidis, ``Fedpoirec: Privacy-preserving federated poi recommendation with social influence,'' \emph{Information Sciences}, vol. 623, pp. 767--790, 2023.

\bibitem{wu2022fedattack}
C.~Wu, F.~Wu, T.~Qi, Y.~Huang, and X.~Xie, ``Fedattack: Effective and covert poisoning attack on federated recommendation via hard sampling,'' in \emph{SIGKDD}, 2022, pp. 4164--4172.

\bibitem{yu2023untargeted}
Y.~Yu, Q.~Liu, L.~Wu, R.~Yu, S.~L. Yu, and Z.~Zhang, ``Untargeted attack against federated recommendation systems via poisonous item embeddings and the defense,'' in \emph{AAAI}, 2023, pp. 4854--4863.

\bibitem{zhang2022pipattack}
S.~Zhang, H.~Yin, T.~Chen, Z.~Huang, Q.~V.~H. Nguyen, and L.~Cui, ``Pipattack: Poisoning federated recommender systems for manipulating item promotion,'' in \emph{WSDM}, 2022, pp. 1415--1423.

\bibitem{chen2022differential}
C.~Chen, H.~Wu, J.~Su, L.~Lyu, X.~Zheng, and L.~Wang, ``Differential private knowledge transfer for privacy-preserving cross-domain recommendation,'' in \emph{WWW}, 2022, pp. 1455--1465.

\bibitem{zhao2016survey}
S.~Zhao, I.~King, and M.~R. Lyu, ``A survey of point-of-interest recommendation in location-based social networks,'' \emph{arXiv preprint arXiv:1607.00647}, 2016.

\bibitem{werneck2020survey}
H.~Werneck, N.~Silva, M.~C. Viana, F.~Mour{\~a}o, A.~C. Pereira, and L.~Rocha, ``A survey on point-of-interest recommendation in location-based social networks,'' in \emph{WebMedia}, 2020, pp. 185--192.

\bibitem{islam2022survey}
M.~A. Islam, M.~M. Mohammad, S.~S.~S. Das, and M.~E. Ali, ``A survey on deep learning based point-of-interest (poi) recommendations,'' \emph{Neurocomputing}, vol. 472, pp. 306--325, 2022.

\bibitem{wang2023survey}
Z.~Wang, W.~H{\"o}pken, and D.~Jannach, ``A survey on point-of-interest recommendations leveraging heterogeneous data,'' \emph{arXiv preprint arXiv:2308.07426}, 2023.

\bibitem{yin2024survey}
H.~Yin, L.~Qu, T.~Chen, W.~Yuan, R.~Zheng, J.~Long, X.~Xia, Y.~Shi, and C.~Zhang, ``On-device recommender systems: A comprehensive survey,'' \emph{arXiv preprint arXiv:2401.11441}, 2024.

\bibitem{yin2013lcars}
H.~Yin, Y.~Sun, B.~Cui, Z.~Hu, and L.~Chen, ``Lcars: a location-content-aware recommender system,'' in \emph{Proceedings of the 19th ACM SIGKDD international conference on Knowledge discovery and data mining}, 2013, pp. 221--229.

\bibitem{yin2015joint}
H.~Yin, X.~Zhou, Y.~Shao, H.~Wang, and S.~Sadiq, ``Joint modeling of user check-in behaviors for point-of-interest recommendation,'' in \emph{CIKM}, 2015, pp. 1631--1640.

\bibitem{wang2015geo}
W.~Wang, H.~Yin, L.~Chen, Y.~Sun, S.~Sadiq, and X.~Zhou, ``Geo-sage: A geographical sparse additive generative model for spatial item recommendation,'' in \emph{SIGKDD}, 2015, pp. 1255--1264.

\bibitem{yin2015modeling}
H.~Yin, B.~Cui, L.~Chen, Z.~Hu, and C.~Zhang, ``Modeling location-based user rating profiles for personalized recommendation,'' \emph{TKDD (TKDD)}, vol.~9, no.~3, pp. 1--41, 2015.

\bibitem{yin2016joint}
H.~Yin, B.~Cui, X.~Zhou, W.~Wang, Z.~Huang, and S.~Sadiq, ``Joint modeling of user check-in behaviors for real-time point-of-interest recommendation,'' \emph{TIS (TOIS)}, vol.~35, no.~2, pp. 1--44, 2016.

\bibitem{wang2016spore}
W.~Wang, H.~Yin, S.~Sadiq, L.~Chen, M.~Xie, and X.~Zhou, ``Spore: A sequential personalized spatial item recommender system,'' in \emph{ICDE}, 2016.

\bibitem{yin2016adapting}
H.~Yin, X.~Zhou, B.~Cui, H.~Wang, K.~Zheng, and Q.~V.~H. Nguyen, ``Adapting to user interest drift for poi recommendation,'' \emph{TKDE}, pp. 2566--2581, 2016.

\bibitem{yin2016discovering}
H.~Yin, Z.~Hu, X.~Zhou, H.~Wang, K.~Zheng, Q.~V.~H. Nguyen, and S.~Sadiq, ``Discovering interpretable geo-social communities for user behavior prediction,'' in \emph{ICDE}, 2016, pp. 942--953.

\bibitem{wang2017location}
H.~Wang, Y.~Fu, Q.~Wang, H.~Yin, C.~Du, and H.~Xiong, ``A location-sentiment-aware recommender system for both home-town and out-of-town users,'' in \emph{Proceedings of the 23rd ACM SIGKDD international conference on knowledge discovery and data mining}, 2017, pp. 1135--1143.

\bibitem{wang2017st}
W.~Wang, H.~Yin, L.~Chen, Y.~Sun, S.~Sadiq, and X.~Zhou, ``St-sage: A spatial-temporal sparse additive generative model for spatial item recommendation,'' \emph{ACM Transactions on Intelligent Systems and Technology (TIST)}, vol.~8, no.~3, pp. 1--25, 2017.

\bibitem{wang2018tpm}
W.~Wang, H.~Yin, X.~Du, Q.~V.~H. Nguyen, and X.~Zhou, ``Tpm: A temporal personalized model for spatial item recommendation,'' \emph{TIST}, 2018.

\bibitem{lian2014geomf}
D.~Lian, C.~Zhao, X.~Xie, G.~Sun, E.~Chen, and Y.~Rui, ``Geomf: joint geographical modeling and matrix factorization for point-of-interest recommendation,'' in \emph{Proceedings of the 20th ACM SIGKDD international conference on Knowledge discovery and data mining}, 2014, pp. 831--840.

\bibitem{lian2018geomf++}
D.~Lian, K.~Zheng, Y.~Ge, L.~Cao, E.~Chen, and X.~Xie, ``Geomf++ scalable location recommendation via joint geographical modeling and matrix factorization,'' \emph{TIS (TOIS)}, vol.~36, no.~3, pp. 1--29, 2018.

\bibitem{rahmani2020lglmf}
H.~A. Rahmani, M.~Aliannejadi, S.~Ahmadian, M.~Baratchi, M.~Afsharchi, and F.~Crestani, ``Lglmf: local geographical based logistic matrix factorization model for poi recommendation,'' in \emph{Information Retrieval Technology: 15th Asia Information Retrieval Societies Conference, AIRS 2019, Hong Kong, China, November 7--9, 2019, Proceedings 15}.\hskip 1em plus 0.5em minus 0.4em\relax Springer, 2020, pp. 66--78.

\bibitem{davtalab2021poi}
M.~Davtalab and A.~A. Alesheikh, ``A poi recommendation approach integrating social spatio-temporal information into probabilistic matrix factorization,'' \emph{Knowledge and Information Systems}, vol.~63, pp. 65--85, 2021.

\bibitem{wang2024sqpmf}
J.~Wang, Z.~Huang, and Z.~Liu, ``Sqpmf: successive point of interest recommendation system based on probability matrix factorization,'' \emph{Applied Intelligence}, vol.~54, no.~1, pp. 680--700, 2024.

\bibitem{zhang2018discrete}
Y.~Zhang, H.~Wang, D.~Lian, I.~W. Tsang, H.~Yin, and G.~Yang, ``Discrete ranking-based matrix factorization with self-paced learning,'' in \emph{SIGKDD}, 2018, pp. 2758--2767.

\bibitem{zhan2019semantic}
G.~Zhan, J.~Xu, Z.~Huang, Q.~Zhang, M.~Xu, and N.~Zheng, ``A semantic sequential correlation based lstm model for next poi recommendation,'' in \emph{2019 20th IEEE International Conference on Mobile Data Management (MDM)}.\hskip 1em plus 0.5em minus 0.4em\relax IEEE, 2019, pp. 128--137.

\bibitem{zhong2021ps}
C.~Zhong, J.~Zhu, and H.~Xi, ``Ps-lstm: Popularity analysis and social network for point-of-interest recommendation in previously unvisited locations,'' in \emph{Proceedings of the 2021 2nd International Conference on Computing, Networks and Internet of Things}, 2021, pp. 1--6.

\bibitem{zhang2022spatio}
Y.~Zhang, P.~Lan, Y.~Wang, and H.~Xiang, ``Spatio-temporal mogrifier lstm and attention network for next poi recommendation,'' in \emph{2022 IEEE International Conference on Web Services (ICWS)}.\hskip 1em plus 0.5em minus 0.4em\relax IEEE, 2022, pp. 17--26.

\bibitem{yang2022attention}
Q.~Yang, S.~Hu, W.~Zhang, and J.~Zhang, ``Attention mechanism and adaptive convolution actuated fusion network for next poi recommendation,'' \emph{International Journal of Intelligent Systems}, vol.~37, no.~10, pp. 7888--7908, 2022.

\bibitem{lai2023poi}
Y.~Lai and X.~Zeng, ``A poi recommendation model for intelligent systems using at-lstm in location-based social network big data,'' \emph{International Journal on Semantic Web and Information Systems (IJSWIS)}, vol.~19, no.~1, pp. 1--15, 2023.

\bibitem{jia2023attention}
Y.~Jia, ``Attention-based time sequence and distance contexts gated recurrent unit for personalized poi recommendation,'' \emph{International Journal of Information Technologies and Systems Approach (IJITSA)}, vol.~16, no.~2, pp. 1--14, 2023.

\bibitem{wu2024dlan}
Y.~Wu, X.~Jiao, Q.~Hao, Y.~Xiao, and W.~Zheng, ``Dlan: Modeling user long-and short-term preferences based on double-layer attention network for next point-of-interest recommendation,'' \emph{Journal of Intelligent \& Fuzzy Systems}, no. Preprint, pp. 1--15, 2024.

\bibitem{halder2021transformer}
S.~Halder, K.~H. Lim, J.~Chan, and X.~Zhang, ``Transformer-based multi-task learning for queuing time aware next poi recommendation,'' in \emph{Pacific-Asia Conference on Knowledge Discovery and Data Mining}.\hskip 1em plus 0.5em minus 0.4em\relax Springer, 2021, pp. 510--523.

\bibitem{yang2022getnext}
S.~Yang, J.~Liu, and K.~Zhao, ``Getnext: trajectory flow map enhanced transformer for next poi recommendation,'' in \emph{Proceedings of the 45th International ACM SIGIR Conference on research and development in information retrieval}, 2022, pp. 1144--1153.

\bibitem{hossain2022caran}
M.~B. Hossain, M.~S. Arefin, I.~H. Sarker, M.~Kowsher, P.~K. Dhar, and T.~Koshiba, ``Caran: A context-aware recency-based attention network for point-of-interest recommendation,'' \emph{IEEE Access}, vol.~10, pp. 36\,299--36\,310, 2022.

\bibitem{zhong2022joint}
H.~Zhong, W.~He, L.~Cui, L.~Liu, Z.~Yan, and K.~Zhao, ``Joint attention networks with inherent and contextual preference-awareness for successive poi recommendation,'' \emph{Data Science and Engineering}, vol.~7, no.~4, pp. 370--382, 2022.

\bibitem{li2022using}
R.~Li, J.~Guo, C.~Liu, Z.~Li, and S.~Zhang, ``Using attributes explicitly reflecting user preference in a self-attention network for next poi recommendation,'' \emph{ISPRS International Journal of Geo-Information}, vol.~11, no.~8, p. 440, 2022.

\bibitem{li2022spatio}
Z.~Li, X.~Huang, C.~Liu, and W.~Yang, ``Spatio-temporal unequal interval correlation-aware self-attention network for next poi recommendation,'' \emph{ISPRS International Journal of Geo-Information}, vol.~11, no.~11, p. 543, 2022.

\bibitem{zang2021cha}
H.~Zang, D.~Han, X.~Li, Z.~Wan, and M.~Wang, ``Cha: Categorical hierarchy-based attention for next poi recommendation,'' \emph{TIS (TOIS)}, vol.~40, no.~1, pp. 1--22, 2021.

\bibitem{wang2022next}
X.~Wang, Y.~Liu, X.~Zhou, Z.~Leng, and X.~Wang, ``Next poi recommendation method based on category preference and attention mechanism in lbsns,'' in \emph{Asia-Pacific Web (APWeb) and Web-Age Information Management (WAIM) Joint International Conference on Web and Big Data}.\hskip 1em plus 0.5em minus 0.4em\relax Springer, 2022, pp. 12--19.

\bibitem{li2022multi}
H.~Li, P.~Yue, S.~Li, F.~Yu, C.~Zhang, C.~Yang, and L.~Jiang, ``Multi-view self-attention network for next poi recommendation,'' in \emph{2022 IEEE Smartworld, Ubiquitous Intelligence \& Computing, Scalable Computing \& Communications, Digital Twin, Privacy Computing, Metaverse, Autonomous \& Trusted Vehicles (SmartWorld/UIC/ScalCom/DigitalTwin/PriComp/Meta)}.\hskip 1em plus 0.5em minus 0.4em\relax IEEE, 2022, pp. 1825--1832.

\bibitem{wu2023reason}
Y.~Wu, G.~Zhao, M.~Li, Z.~Zhang, and X.~Qian, ``Reason generation for point of interest recommendation via a hierarchical attention-based transformer model,'' \emph{IEEE Transactions on Multimedia}, 2023.

\bibitem{xie2023hierarchical}
J.~Xie and Z.~Chen, ``Hierarchical transformer with spatio-temporal context aggregation for next point-of-interest recommendation,'' \emph{TIS}, vol.~42, no.~2, pp. 1--30, 2023.

\bibitem{halder2023capacity}
S.~Halder, K.~H. Lim, J.~Chan, and X.~Zhang, ``Capacity-aware fair poi recommendation combining transformer neural networks and resource allocation policy,'' \emph{Applied Soft Computing}, vol. 147, p. 110720, 2023.

\bibitem{he2023feature}
Y.~He, W.~Zhou, F.~Luo, M.~Gao, and J.~Wen, ``Feature-based poi grouping with transformer for next point of interest recommendation,'' \emph{Applied Soft Computing}, vol. 147, p. 110754, 2023.

\bibitem{jiang2023temporal}
S.~Jiang and J.~Wu, ``Temporal-geographical attention-based transformer for point-of-interest recommendation,'' \emph{Journal of Intelligent \& Fuzzy Systems}, no. Preprint, pp. 1--11, 2023.

\bibitem{xu2023revisiting}
X.~Xu, T.~Suzumura, J.~Yong, M.~Hanai, C.~Yang, H.~Kanezashi, R.~Jiang, and S.~Fukushima, ``Revisiting mobility modeling with graph: A graph transformer model for next point-of-interest recommendation,'' in \emph{Proceedings of the 31st ACM International Conference on Advances in Geographic Information Systems}, 2023, pp. 1--10.

\bibitem{ho2022poibert}
N.~L. Ho and K.~H. Lim, ``Poibert: A transformer-based model for the tour recommendation problem,'' in \emph{2022 IEEE International Conference on Big Data (Big Data)}.\hskip 1em plus 0.5em minus 0.4em\relax IEEE, 2022, pp. 5925--5933.

\bibitem{qin2022next}
Y.~Qin, Y.~Fang, H.~Luo, F.~Zhao, and C.~Wang, ``Next point-of-interest recommendation with auto-correlation enhanced multi-modal transformer network,'' in \emph{Proceedings of the 45th International ACM SIGIR Conference on Research and Development in Information Retrieval}, 2022, pp. 2612--2616.

\bibitem{wang2023context}
D.~Wang, F.~Wan, D.~Yu, Y.~Shen, Z.~Xiang, and Y.~Xu, ``Context-and category-aware double self-attention model for next poi recommendation,'' \emph{Applied Intelligence}, vol.~53, no.~15, pp. 18\,355--18\,380, 2023.

\bibitem{liu2023poi}
X.~Liu, ``Poi recommendation model using multi-head attention in location-based social network big data,'' \emph{International Journal of Information Technologies and Systems Approach (IJITSA)}, vol.~16, no.~2, pp. 1--16, 2023.

\bibitem{cao2023improving}
G.~Cao, S.~Cui, and I.~Joe, ``Improving the spatial--temporal aware attention network with dynamic trajectory graph learning for next point-of-interest recommendation,'' \emph{Information Processing \& Management}, vol.~60, no.~3, p. 103335, 2023.

\bibitem{wu2023muti}
Y.~Wu, X.~Jin, and H.~Huang, ``Muti-channel graph attention networks for poi recommendation,'' \emph{Journal of Intelligent \& Fuzzy Systems}, vol.~44, no.~5, pp. 8375--8385, 2023.

\bibitem{ou2023sta}
J.~Ou, H.~Jin, X.~Wang, H.~Jiang, X.~Wang, and C.~Zhou, ``Sta-tcn: Spatial-temporal attention over temporal convolutional network for next point-of-interest recommendation,'' \emph{TKDD}, vol.~17, no.~9, pp. 1--19, 2023.

\bibitem{xia2023effective}
Q.~Xia, T.~Hara, T.~Maekawa, M.~Kurokawa, and K.~Yonekawa, ``An effective and efficient self-attention based model for next poi recommendation,'' in \emph{2023 IEEE International Conference on Pervasive Computing and Communications Workshops and other Affiliated Events (PerCom Workshops)}.\hskip 1em plus 0.5em minus 0.4em\relax IEEE, 2023, pp. 478--483.

\bibitem{xu2023spatio}
S.~Xu, Q.~Huang, and Z.~Zou, ``Spatio-temporal transformer recommender: next location recommendation with attention mechanism by mining the spatio-temporal relationship between visited locations,'' \emph{ISPRS International Journal of Geo-Information}, vol.~12, no.~2, p.~79, 2023.

\bibitem{xia2023bayes}
J.~Xia, Y.~Yang, S.~Wang, H.~Yin, J.~Cao, and S.~Y. Philip, ``Bayes-enhanced multi-view attention networks for robust poi recommendation,'' \emph{TKDE}, 2023.

\bibitem{kumar2024modified}
A.~Kumar, D.~K. Jain, A.~Mallik, and S.~Kumar, ``Modified node2vec and attention based fusion framework for next poi recommendation,'' \emph{Information Fusion}, vol. 101, p. 101998, 2024.

\bibitem{he2024imnext}
X.~He, W.~He, Y.~Liu, X.~Lu, Y.~Xiao, and Y.~Liu, ``Imnext: Irregular interval attention and multi-task learning for next poi recommendation,'' \emph{Knowledge-Based Systems}, vol. 293, p. 111674, 2024.

\bibitem{cheng2024point}
S.~Cheng, Z.~Wu, M.~Qian, and W.~Huang, ``Point-of-interest recommendation based on bidirectional self-attention mechanism by fusing spatio-temporal preference,'' \emph{Multimedia Tools and Applications}, vol.~83, no.~9, pp. 26\,333--26\,347, 2024.

\bibitem{wu2024reason}
Y.~Wu, G.~Zhao, M.~Li, Z.~Zhang, and X.~Qian, ``Reason generation for point of interest recommendation via a hierarchical attention-based transformer model,'' \emph{IEEE Transactions on Multimedia}, 2024.

\bibitem{feng2024rotan}
S.~Feng, F.~Meng, L.~Chen, S.~Shang, and Y.~S. Ong, ``Rotan: A rotation-based temporal attention network for time-specific next poi recommendation,'' in \emph{Proceedings of the 30th ACM SIGKDD Conference on Knowledge Discovery and Data Mining}.\hskip 1em plus 0.5em minus 0.4em\relax New York, NY, USA: Association for Computing Machinery, 2024, p. 759–770.

\bibitem{han2020stgcn}
H.~Han, M.~Zhang, M.~Hou, F.~Zhang, Z.~Wang, E.~Chen, H.~Wang, J.~Ma, and Q.~Liu, ``Stgcn: a spatial-temporal aware graph learning method for poi recommendation,'' in \emph{2020 IEEE International Conference on Data Mining (ICDM)}.\hskip 1em plus 0.5em minus 0.4em\relax IEEE, 2020, pp. 1052--1057.

\bibitem{zhang2021leveraging}
J.~Zhang, X.~Liu, X.~Zhou, and X.~Chu, ``Leveraging graph neural networks for point-of-interest recommendations,'' \emph{Neurocomputing}, vol. 462, pp. 1--13, 2021.

\bibitem{mo2022gn}
F.~Mo and H.~Yamana, ``Gn-gcn: combining geographical neighbor concept with graph convolution network for poi recommendation,'' in \emph{International Conference on Information Integration and Web}.\hskip 1em plus 0.5em minus 0.4em\relax Springer, 2022, pp. 153--165.

\bibitem{li2022attention}
Q.~Li, X.~Xu, X.~Liu, and Q.~Chen, ``An attention-based spatiotemporal ggnn for next poi recommendation,'' \emph{IEEE Access}, vol.~10, pp. 26\,471--26\,480, 2022.

\bibitem{zhao2023poi}
Y.~Zhao, Y.~Zhang, D.~Amagata, and T.~Hara, ``Poi recommendation by learning short-, long-and mid-term preferences through gnn,'' in \emph{2023 24th IEEE International Conference on Mobile Data Management (MDM)}.\hskip 1em plus 0.5em minus 0.4em\relax IEEE, 2023, pp. 1--10.

\bibitem{zhang2024hybrid}
J.~Zhang and W.~Ma, ``Hybrid structural graph attention network for poi recommendation,'' \emph{Expert Systems with Applications}, vol. 248, p. 123436, 2024.

\bibitem{liu2024poi}
Z.~Liu, L.~Meng, Q.~Z. Sheng, D.~Chu, J.~Yu, and X.~Song, ``Poi recommendation for random groups based on cooperative graph neural networks,'' \emph{Information Processing \& Management}, vol.~61, no.~3, p. 103676, 2024.

\bibitem{zhang2024hyper}
J.~Zhang, Y.~Li, R.~Zou, J.~Zhang, R.~Jiang, Z.~Fan, and X.~Song, ``Hyper-relational knowledge graph neural network for next poi recommendation,'' \emph{World Wide Web}, vol.~27, no.~4, pp. 1--19, 2024.

\bibitem{zhou2021self}
F.~Zhou, Y.~Dai, Q.~Gao, P.~Wang, and T.~Zhong, ``Self-supervised human mobility learning for next location prediction and trajectory classification,'' \emph{Knowledge-Based Systems}, vol. 228, p. 107214, 2021.

\bibitem{li2022self}
Y.~Li, T.~Chen, P.-F. Zhang, Z.~Huang, and H.~Yin, ``Self-supervised graph-based point-of-interest recommendation,'' \emph{arXiv preprint arXiv:2210.12506}, 2022.

\bibitem{liu2023self}
J.~Liu, H.~Gao, C.~Shi, H.~Cheng, and Q.~Xie, ``Self-supervised spatio-temporal graph learning for point-of-interest recommendation,'' \emph{Applied Sciences}, vol.~13, no.~15, p. 8885, 2023.

\bibitem{gao2022self}
Q.~Gao, W.~Wang, K.~Zhang, X.~Yang, C.~Miao, and T.~Li, ``Self-supervised representation learning for trip recommendation,'' \emph{Knowledge-Based Systems}, vol. 247, p. 108791, 2022.

\bibitem{wang2023exploring}
D.~Wang, C.~Chen, C.~Di, and M.~Shu, ``Exploring behavior patterns for next-poi recommendation via graph self-supervised learning,'' \emph{Electronics}, vol.~12, no.~8, p. 1939, 2023.

\bibitem{jiang2023modeling}
S.~Jiang, W.~He, L.~Cui, Y.~Xu, and L.~Liu, ``Modeling long-and short-term user preferences via self-supervised learning for next poi recommendation,'' \emph{TKDD}, vol.~17, no.~9, pp. 1--20, 2023.

\bibitem{gao2023predicting}
Q.~Gao, J.~Hong, X.~Xu, P.~Kuang, F.~Zhou, and G.~Trajcevski, ``Predicting human mobility via self-supervised disentanglement learning,'' \emph{TKDE}, 2023.

\bibitem{wang2024graph}
C.~Wang, O.~Tsepa, J.~Ma, and B.~Wang, ``Graph-mamba: Towards long-range graph sequence modeling with selective state spaces,'' \emph{arXiv preprint arXiv:2402.00789}, 2024.

\bibitem{fu2024contrastive}
J.~Fu, R.~Gao, Y.~Yu, J.~Wu, J.~Li, D.~Liu, and Z.~Ye, ``Contrastive graph learning long and short-term interests for poi recommendation,'' \emph{Expert Systems with Applications}, vol. 238, p. 121931, 2024.

\bibitem{zhou2024cllp}
H.~Zhou, Z.~Jia, H.~Zhu, and Z.~Zhang, ``Cllp: Contrastive learning framework based on latent preferences for next poi recommendation,'' in \emph{ACM SIGIR}, 2024, pp. 1473--1482.

\bibitem{feng2024move}
S.~Feng, H.~Lyu, F.~Li, Z.~Sun, and C.~Chen, ``Where to move next: Zero-shot generalization of llms for next poi recommendation,'' in \emph{2024 IEEE Conference on Artificial Intelligence (CAI)}.\hskip 1em plus 0.5em minus 0.4em\relax IEEE, 2024, pp. 1530--1535.

\bibitem{li2024large}
P.~Li, M.~de~Rijke, H.~Xue, S.~Ao, Y.~Song, and F.~D. Salim, ``Large language models for next point-of-interest recommendation,'' \emph{arXiv preprint arXiv:2404.17591}, 2024.

\bibitem{qin2023diffusion}
Y.~Qin, H.~Wu, W.~Ju, X.~Luo, and M.~Zhang, ``A diffusion model for poi recommendation,'' \emph{TIS}, vol.~42, no.~2, pp. 1--27, 2023.

\bibitem{wang2024dsdrec}
Z.~Wang, J.~Zeng, L.~Zhong, L.~Liu, M.~Gao, and J.~Wen, ``Dsdrec: Next poi recommendation using deep semantic extraction and diffusion model,'' \emph{Information Sciences}, p. 121004, 2024.

\bibitem{zuo2024diff}
J.~Zuo and Y.~Zhang, ``Diff-dgmn: A diffusion-based dual graph multi-attention network for poi recommendation,'' \emph{IEEE Internet of Things Journal}, pp. 1--1, 2024.

\bibitem{long2024diffusion}
J.~Long, G.~Ye, T.~Chen, Y.~Wang, M.~Wang, and H.~Yin, ``Diffusion-based cloud-edge-device collaborative learning for next poi recommendations,'' in \emph{KDD}, 2024, pp. 2026--2036.

\bibitem{zhou2023uncertainty}
F.~Zhou, T.~Qian, Y.~Mo, Z.~Cheng, C.~Xiao, J.~Wu, and G.~Trajcevski, ``Uncertainty-aware heterogeneous representation learning in poi recommender systems,'' \emph{IEEE T SYST MAN CY-S}, pp. 4522--4535, 2023.

\bibitem{li2022linking}
Y.~Li, Y.~Sang, W.~Chen, and L.~Zhao, ``Linking check-in data to users on location-aware social networks,'' in \emph{Pacific Rim International Conference on Artificial Intelligence}.\hskip 1em plus 0.5em minus 0.4em\relax Springer, 2022, pp. 489--503.

\bibitem{wang2018exploiting}
H.~Wang, H.~Shen, W.~Ouyang, and X.~Cheng, ``Exploiting poi-specific geographical influence for point-of-interest recommendation,'' in \emph{IJCAI}, 2018, pp. 3877--3883.

\bibitem{yin2017spatial}
H.~Yin, W.~Wang, H.~Wang, L.~Chen, and X.~Zhou, ``Spatial-aware hierarchical collaborative deep learning for poi recommendation,'' \emph{TKDE}, vol.~29, no.~11, pp. 2537--2551, 2017.

\bibitem{zhao2020go}
P.~Zhao, A.~Luo, Y.~Liu, J.~Xu, Z.~Li, F.~Zhuang, V.~S. Sheng, and X.~Zhou, ``Where to go next: A spatio-temporal gated network for next poi recommendation,'' \emph{TKDE}, vol.~34, no.~5, pp. 2512--2524, 2020.

\bibitem{wang2023adaptive}
Z.~Wang, Y.~Zhu, C.~Wang, W.~Ma, B.~Li, and J.~Yu, ``Adaptive graph representation learning for next poi recommendation,'' in \emph{Proceedings of the 46th International ACM SIGIR Conference on Research and Development in Information Retrieval}, 2023, pp. 393--402.

\bibitem{long2023model}
J.~Long, T.~Chen, Q.~V.~H. Nguyen, G.~Xu, K.~Zheng, and H.~Yin, ``Model-agnostic decentralized collaborative learning for on-device poi recommendation,'' in \emph{Proceedings of the 46th International ACM SIGIR Conference on Research and Development in Information Retrieval}, 2023, pp. 423--432.

\bibitem{zheng2024decentralized}
R.~Zheng, L.~Qu, T.~Chen, L.~Cui, Y.~Shi, and H.~Yin, ``Decentralized collaborative learning with adaptive reference data for on-device poi recommendation,'' in \emph{WWW}, 2024, pp. 3930--3939.

\bibitem{wang2020next}
Q.~Wang, H.~Yin, T.~Chen, Z.~Huang, H.~Wang, Y.~Zhao, and N.~Q. Viet~Hung, ``Next point-of-interest recommendation on resource-constrained mobile devices,'' in \emph{WWW}, 2020, pp. 906--916.

\bibitem{wang2021poi}
L.-e. Wang, Y.~Wang, Y.~Bai, P.~Liu, and X.~Li, ``Poi recommendation with federated learning and privacy preserving in cross domain recommendation,'' in \emph{INFOCOM WKSHPS}.\hskip 1em plus 0.5em minus 0.4em\relax IEEE, 2021, pp. 1--6.

\bibitem{huang2022geographical}
J.~Huang, Z.~Tong, and Z.~Feng, ``Geographical poi recommendation for internet of things: A federated learning approach using matrix factorization,'' \emph{IJCS}, p. e5161, 2022.

\bibitem{ye2023adaptive}
Z.~Ye, X.~Zhang, X.~Chen, H.~Xiong, and D.~Yu, ``Adaptive clustering based personalized federated learning framework for next poi recommendation with location noise,'' \emph{IEEE TKDE}, 2023.

\bibitem{zhong2024scfl}
L.~Zhong, J.~Zeng, Z.~Wang, W.~Zhou, and J.~Wen, ``Scfl: Spatio-temporal consistency federated learning for next poi recommendation,'' \emph{IPM}, vol.~61, no.~6, p. 103852, 2024.

\bibitem{dong2024sfl}
X.~Dong, J.~Zeng, J.~Wen, M.~Gao, and W.~Zhou, ``Sfl: A semantic-based federated learning method for poi recommendation,'' \emph{Information Sciences}, p. 121057, 2024.

\bibitem{zhang2023fine}
X.~Zhang, Z.~Ye, J.~Lu, F.~Zhuang, Y.~Zheng, and D.~Yu, ``Fine-grained preference-aware personalized federated poi recommendation with data sparsity,'' in \emph{Proceedings of the 46th International ACM SIGIR Conference on Research and Development in Information Retrieval}, 2023, pp. 413--422.

\bibitem{guo2021prefer}
Y.~Guo, F.~Liu, Z.~Cai, H.~Zeng, L.~Chen, T.~Zhou, and N.~Xiao, ``Prefer: Point-of-interest recommendation with efficiency and privacy-preservation via federated edge learning,'' \emph{Proceedings of the ACM on Interactive, Mobile, Wearable and Ubiquitous Technologies}, vol.~5, no.~1, pp. 1--25, 2021.

\bibitem{dong2023sequential}
Q.~Dong, B.~Liu, X.~Zhang, J.~Qin, and B.~Wang, ``Sequential poi recommend based on personalized federated learning,'' \emph{Neural Processing Letters}, vol.~55, no.~6, pp. 7351--7368, 2023.

\bibitem{chen2018privacy}
C.~Chen, Z.~Liu, P.~Zhao, J.~Zhou, and X.~Li, ``Privacy preserving point-of-interest recommendation using decentralized matrix factorization,'' in \emph{AAAI}, 2018, pp. 257--264.

\bibitem{an2024nrdl}
J.~An, G.~Li, and W.~Jiang, ``{NRDL}: Decentralized user preference learning for privacy-preserving next poi recommendation,'' \emph{Expert Systems with Applications}, vol. 239, p. 122421, 2024.

\bibitem{rong2022fedrecattack}
D.~Rong, S.~Ye, R.~Zhao, H.~N. Yuen, J.~Chen, and Q.~He, ``Fedrecattack: model poisoning attack to federated recommendation,'' in \emph{ICDE}, 2022, pp. 2643--2655.

\bibitem{wang2018ghost}
G.~Wang, B.~Wang, T.~Wang, A.~Nika, H.~Zheng, and B.~Y. Zhao, ``Ghost riders: Sybil attacks on crowdsourced mobile mapping services,'' \emph{IEEE/ACM transactions on networking}, vol.~26, no.~3, pp. 1123--1136, 2018.

\bibitem{zhang2022loki}
H.~Zhang, Y.~Li, B.~Ding, and J.~Gao, ``Loki: a practical data poisoning attack framework against next item recommendations,'' \emph{TKDE}, vol.~35, no.~5, pp. 5047--5059, 2022.

\bibitem{yuan2023manipulating}
W.~Yuan, Q.~V.~H. Nguyen, T.~He, L.~Chen, and H.~Yin, ``Manipulating federated recommender systems: Poisoning with synthetic users and its countermeasures,'' \emph{SIGIR}, pp. 1690--1699, 2023.

\bibitem{long2024physical}
J.~Long, T.~Chen, G.~Ye, K.~Zheng, Q.~V.~H. Nguyen, and H.~Yin, ``Physical trajectory inference attack and defense in decentralized poi recommendation,'' in \emph{Proceedings of the ACM on Web Conference 2024}, 2024, pp. 3379--3387.

\bibitem{kuang2020providing}
L.~Kuang, S.~Tu, Y.~Zhang, and X.~Yang, ``Providing privacy preserving in next poi recommendation for mobile edge computing,'' \emph{Journal of Cloud Computing}, vol.~9, pp. 1--11, 2020.

\bibitem{chen2020practical}
C.~Chen, J.~Zhou, B.~Wu, W.~Fang, L.~Wang, Y.~Qi, and X.~Zheng, ``Practical privacy preserving poi recommendation,'' \emph{ACM TIST}, vol.~11, no.~5, pp. 1--20, 2020.

\bibitem{liu2017privacy}
A.~Liu, W.~Wang, Z.~Li, G.~Liu, Q.~Li, X.~Zhou, and X.~Zhang, ``A privacy-preserving framework for trust-oriented point-of-interest recommendation,'' \emph{IEEE Access}, vol.~6, pp. 393--404, 2017.

\bibitem{huo2021privacy}
Y.~Huo, B.~Chen, J.~Tang, and Y.~Zeng, ``Privacy-preserving point-of-interest recommendation based on geographical and social influence,'' \emph{Information Sciences}, vol. 543, pp. 202--218, 2021.

\bibitem{li2018survey}
Y.~Li, M.~Yang, and Z.~M. Zhang, ``A survey of multi-view representation learning,'' \emph{TKDE}, vol.~31, no.~10, pp. 1863--1883, 2018.

\bibitem{han2020survey}
K.~Han, Y.~Wang, H.~Chen, X.~Chen, J.~Guo, Z.~Liu, Y.~Tang, A.~Xiao, C.~Xu, Y.~Xu \emph{et~al.}, ``A survey on visual transformer,'' \emph{arXiv preprint arXiv:2012.12556}, 2020.

\bibitem{li2023bilstm}
A.~Li and F.~Liu, ``A bilstm-attention-based point-of-interest recommendation algorithm,'' \emph{Journal of Intelligent Systems}, vol.~32, no.~1, p. 20230033, 2023.

\bibitem{munro1986self}
P.~Munro, ``Self-supervised learning: A scheme for discovery of" natural" categories by single units,'' in \emph{Proceedings of the Annual Meeting of the Cognitive Science Society}, vol.~8, 1986.

\bibitem{yu2023self}
J.~Yu, H.~Yin, X.~Xia, T.~Chen, J.~Li, and Z.~Huang, ``Self-supervised learning for recommender systems: A survey,'' \emph{TKDE}, vol.~36, no.~1, pp. 335--355, 2023.

\bibitem{kingma2013auto}
D.~P. Kingma, ``Auto-encoding variational bayes,'' \emph{arXiv preprint arXiv:1312.6114}, 2013.

\bibitem{kingma2019introduction}
D.~P. Kingma, M.~Welling \emph{et~al.}, ``An introduction to variational autoencoders,'' \emph{Foundations and Trends{\textregistered} in Machine Learning}, vol.~12, no.~4, pp. 307--392, 2019.

\bibitem{zhao2018go}
P.~Zhao, H.~Zhu, Y.~Liu, Z.~Li, J.~Xu, and V.~S. Sheng, ``Where to go next: A spatio-temporal lstm model for next poi recommendation,'' \emph{arXiv preprint arXiv:1806.06671}, 2018.

\bibitem{mcmahan2017communication}
B.~McMahan, E.~Moore, D.~Ramage, S.~Hampson, and B.~A. y~Arcas, ``Communication-efficient learning of deep networks from decentralized data,'' in \emph{Artificial intelligence and statistics}.\hskip 1em plus 0.5em minus 0.4em\relax PMLR, 2017, pp. 1273--1282.

\bibitem{xie2016learning}
M.~Xie, H.~Yin, H.~Wang, F.~Xu, W.~Chen, and S.~Wang, ``Learning graph-based poi embedding for location-based recommendation,'' in \emph{CIKM}, 2016.

\bibitem{sadek2012svd}
R.~A. Sadek, ``Svd based image processing applications: state of the art, contributions and research challenges,'' \emph{arXiv preprint arXiv:1211.7102}, 2012.

\bibitem{brown2015methods}
S.~G. Brown, S.~Eberly, P.~Paatero, and G.~A. Norris, ``Methods for estimating uncertainty in pmf solutions: Examples with ambient air and water quality data and guidance on reporting pmf results,'' \emph{Science of the Total Environment}, vol. 518, pp. 626--635, 2015.

\bibitem{song2013hierarchical}
H.~A. Song and S.-Y. Lee, ``Hierarchical representation using nmf,'' in \emph{Neural Information Processing: 20th International Conference, ICONIP 2013, Daegu, Korea, November 3-7, 2013. Proceedings, Part I 20}.\hskip 1em plus 0.5em minus 0.4em\relax Springer, 2013, pp. 466--473.

\bibitem{schuster1997bidirectional}
M.~Schuster and K.~K. Paliwal, ``Bidirectional recurrent neural networks,'' \emph{IEEE TSP}, pp. 2673--2681, 1997.

\bibitem{scarselli2008graph}
F.~Scarselli, M.~Gori, A.~C. Tsoi, M.~Hagenbuchner, and G.~Monfardini, ``The graph neural network model,'' \emph{IEEE transactions on neural networks}, vol.~20, no.~1, pp. 61--80, 2008.

\bibitem{xu2018powerful}
K.~Xu, W.~Hu, J.~Leskovec, and S.~Jegelka, ``How powerful are graph neural networks?'' \emph{arXiv preprint arXiv:1810.00826}, 2018.

\bibitem{liu2021self}
X.~Liu, F.~Zhang, Z.~Hou, L.~Mian, Z.~Wang, J.~Zhang, and J.~Tang, ``Self-supervised learning: Generative or contrastive,'' \emph{TKDE}, vol.~35, no.~1, pp. 857--876, 2021.

\bibitem{chen2020simple}
T.~Chen, S.~Kornblith, M.~Norouzi, and G.~Hinton, ``A simple framework for contrastive learning of visual representations,'' in \emph{International conference on machine learning}.\hskip 1em plus 0.5em minus 0.4em\relax PMLR, 2020, pp. 1597--1607.

\bibitem{gui2024survey}
J.~Gui, T.~Chen, J.~Zhang, Q.~Cao, Z.~Sun, H.~Luo, and D.~Tao, ``A survey on self-supervised learning: Algorithms, applications, and future trends,'' \emph{TPAMI}, 2024.

\bibitem{devlin2019bert}
J.~Devlin, M.-W. Chang, K.~Lee, and K.~Toutanova, ``Bert: Pre-training of deep bidirectional transformers for language understanding,'' in \emph{Proceedings of the 2019 Conference of the North American Chapter of the Association for Computational Linguistics: Human Language Technologies}, 2019, pp. 4171--4186.

\bibitem{radford2019language}
A.~Radford, J.~Wu, R.~Child, D.~Luan, D.~Amodei, I.~Sutskever \emph{et~al.}, ``Language models are unsupervised multitask learners,'' \emph{OpenAI blog}, vol.~1, no.~8, p.~9, 2019.

\bibitem{raffel2020exploring}
C.~Raffel, N.~Shazeer, A.~Roberts, K.~Lee, S.~Narang, M.~Matena, Y.~Zhou, W.~Li, and P.~J. Liu, ``Exploring the limits of transfer learning with a unified text-to-text transformer,'' \emph{Journal of machine learning research}, vol.~21, no. 140, pp. 1--67, 2020.

\bibitem{sohl2015deep}
J.~Sohl-Dickstein, E.~Weiss, N.~Maheswaranathan, and S.~Ganguli, ``Deep unsupervised learning using nonequilibrium thermodynamics,'' in \emph{International conference on machine learning}.\hskip 1em plus 0.5em minus 0.4em\relax PMLR, 2015, pp. 2256--2265.

\bibitem{chen2025self}
L.~Chen and G.~Zhu, ``Self-supervised contrastive learning for itinerary recommendation,'' \emph{Expert Systems with Applications}, vol. 268, p. 126246, 2025.

\bibitem{xi2023rise}
Z.~Xi, W.~Chen, X.~Guo, W.~He, Y.~Ding, B.~Hong, M.~Zhang, J.~Wang, S.~Jin, E.~Zhou \emph{et~al.}, ``The rise and potential of large language model based agents: A survey,'' \emph{Science China Information Sciences}, p. 121101, 2025.

\bibitem{liu2024semantic}
Y.~Liu, C.~Kuai, H.~Ma, X.~Liao, B.~Y. He, and J.~Ma, ``Semantic trajectory data mining with llm-informed poi classification,'' \emph{arXiv preprint arXiv:2405.11715}, 2024.

\bibitem{ning2024cheatagent}
L.-b. Ning, S.~Wang, W.~Fan, Q.~Li, X.~Xu, H.~Chen, and F.~Huang, ``Cheatagent: Attacking llm-empowered recommender systems via llm agent,'' in \emph{Proceedings of the 30th ACM SIGKDD Conference on Knowledge Discovery and Data Mining}, 2024, pp. 2284--2295.

\bibitem{li2024survey}
Y.~Li, X.~Feng, C.~Chen, and Q.~Yang, ``A survey on recommendation unlearning: Fundamentals, taxonomy, evaluation, and open questions,'' \emph{arXiv preprint arXiv:2412.12836}, 2024.

\bibitem{costan2016intel}
V.~Costan, ``Intel sgx explained,'' \emph{IACR Cryptol, EPrint Arch}, 2016.

\bibitem{ngabonziza2016trustzone}
B.~Ngabonziza, D.~Martin, A.~Bailey, H.~Cho, and S.~Martin, ``Trustzone explained: Architectural features and use cases,'' in \emph{IEEE CIC}.\hskip 1em plus 0.5em minus 0.4em\relax IEEE, 2016, pp. 445--451.

\end{thebibliography}

\end{multicols}
\vspace{-0.25in}

\end{document}